\documentclass[9pt]{article}

\usepackage[margin=0.9in]{geometry}
\usepackage[english]{babel}
\usepackage[utf8]{inputenc}
\usepackage{graphicx}
\usepackage{caption}
\usepackage{subcaption}
\usepackage{soul}
\usepackage{float}
\usepackage{amsmath}
\usepackage{url}
\usepackage{comment}
\usepackage[table]{xcolor}
\usepackage{array}
\usepackage{multirow}
\usepackage{diagbox}

\usepackage{tikz}
\usetikzlibrary{shapes}

\usepackage{authblk}
\usepackage[nottoc]{tocbibind}
\usepackage{mathrsfs}

\usepackage{algorithm}
\usepackage{algpseudocode}

\renewcommand{\arraystretch}{1.35}
\definecolor{figblue}{RGB}{31,119,180}
\definecolor{figorange}{RGB}{255,127,14}
\definecolor{figgreen}{RGB}{44,160,44}

\definecolor{ForestGreen}{HTML}{228b22}

\title{Statistical Laws of Human Mobility under Large-Scale Behavioral Shock}

\author[1,2]{Alberto Amaduzzi}
\author[2]{Lucila G. Alvarez-Zuzek \thanks{lalvarezzuzek@fbk.eu}}
\author[3,4,5]{Valeria d'Andrea}
\author[6]{Simone Centellegher}
\author[7]{Lorenzo Lucchini}
\author[8]{Brennan Lake}
\author[3,4,5]{Manlio De Domenico}
\author[6]{Bruno Lepri}
\author[9]{Marta C. Gonzalez}
\author[1]{Armando Bazzani}
\author[2]{Riccardo Gallotti} %

\affil[1]{Department of Physics, University of Bologna, Bologna, Italy}
\affil[2]{Complex Human Behaviour Laboratory, Fondazione Bruno Kessler, Trento, Italy}
\affil[3]{Department of Physics and Astronomy ``Galileo Galilei'', University of Padua, Via F. Marzolo 8, 35131 Padova, Italy}
\affil[4]{Istituto Nazionale di Fisica Nucleare, Sez. Padova, Padua, Italy}
\affil[5]{Padua Center for Network Medicine, University of Padua, 35131 Padova, Italy}
\affil[6]{Mobile and Social Computing Laboratory, Fondazione Bruno Kessler, Trento, Italy}
\affil[7]{Center for Health Emergencies, Fondazione Bruno Kessler, Trento, Italy}
\affil[8]{Cuebiq Inc., New York, NY, USA}
\affil[9]{Department of Civil and Environmental Engineering, University of California, Berkeley, USA}

\begin{document}

    \maketitle
    
    \begin{abstract}
    Human mobility follows robust statistical laws, but whether these laws persist under large, sudden behavioural shocks remains unclear. Here we analyse privacy-enhanced GPS trajectories of over one million users in California and 180,000 in Massachusetts from January to September 2020, spanning the onset and easing of COVID-19 restrictions. During restrictions, individual ranges contracted below a characteristic scale of about 5~km, while short- and medium-range trip lengths and rare long-distance trips persisted. Exploration slowed from super-diffusive to near-diffusive, and visits concentrated on a few habitual locations. Trajectory shape, once rescaled, was preserved, and the entropy reduction is largely explained by fewer visited locations rather than more regular routines. Recovery after reopening was heterogeneous rather than a uniform return to baseline, and the decline began before mandatory orders. These results suggest that restrictions reorganised, rather than simply rescaled, individual mobility, with implications for mobility-based epidemic models.
    \end{abstract}

    \newpage
    \section*{Introduction}
        \label{introduction}

The COVID-19 pandemic raised fundamental questions about how to contain the spread of a novel disease \cite{LepriCovid,EvaluationCovid} and about the collateral effects of containment strategies \cite{CovidEffectEconomy}. In the absence of vaccines, Non-Pharmaceutical Interventions (NPIs) were the primary tool adopted by governments to limit transmission \cite{NPI1,NPIDD,NPI3}, significantly reshaping human behaviour and mobility patterns \cite{SantanaChanges}. Importantly, a substantial share of these behavioural changes preceded formal mandates, driven by voluntary responses to risk information \cite{Goolsbee2020,Gupta2020,Xiong2020}, a distinction with direct consequences for how mobility-based models should be calibrated. This context produced a large-scale observational perturbation of routine mobility and provided the scientific community with new challenges to model epidemic spread under changing behavioural regimes \cite{model0,model1,model2,model3,model4}.

Aggregate and anonymised mobile phone data have long been shown to support the modelling of epidemic geography \cite{HumanProxiesModelling,MobilityMalaria}, while individual GPS trajectories are seen as a possible resource to improve adaptive estimates of contact patterns and spatial transmission dynamics \cite{Finger2016,LepriCovid,Alessandretti,Chang2021,Kraemer2020}. The COVID-19 pandemic catalysed an unprecedented release of such data, from aggregated public indices \cite{GoogleMobility,AppleMobility} to privacy-enhanced individual-level datasets \cite{Chang2021,lucchini}, enabling analyses at scales previously unavailable. These datasets are nonetheless subject to demographic sampling biases that must be accounted for when drawing population-level inferences \cite{Wesolowski2013,Zhao2016,Li2024}, moreover, the choice of data source and processing pipeline can itself substantially alter the inferred mobility patterns \cite{gallotti2024distorted}. Mobility data are often partial, aggregated, or available at mismatched spatial scales \cite{Gatto2020,Bertuzzo2020,Balcan2009a}, leading studies to complement them with probabilistic mobility models \cite{Aleta2020,Arenas2020,HumanProxiesModelling,CiofidegliAtti2008}. If restrictions alter exploration and visitation patterns in addition to the scale of individuals' movements~\cite{lucchini, lucchini2025socioeconomic,Schlosser2020a}, models that only rescale baseline mobility may fail to reproduce some features of the observed epidemic trajectories.  

The reduction in mobility was itself heterogeneous: sharp differences have been documented along income gradients \cite{Weill2020,lucchini2025socioeconomic}, urban--rural divides \cite{Li2020}, and partisan lines \cite{Allcott2020,Grossman2020}. Recovery dynamics after restrictions were lifted showed similar heterogeneity across jurisdictions and political contexts \cite{Kim2021,Nguyen2020}. Understanding this variability, and whether standard statistical descriptions of human mobility remain valid under forced restrictions, is the central motivation of this work.

The relationship between trip counts, contacts, and transmission probability did not remain constant across pandemic phases, driven by shifting levels of public awareness \cite{nouvellet,rg_forecast_deaths}, making it challenging to infer viral spread from mobility data alone \cite{Alessandretti}. Motivated by this limitation, in this work we focus on improving the characterisation of  changes in mobility patterns, through geometric measures (radius of gyration, straight line distance), visitation patterns, and information-theoretic complexity, rather than trying to infer transmission patterns from mobility measures.

In this paper, we laverage a large dataset describing individual mobility in the U.S. and we divide the observation period into three phases: pre-restriction
(15 January -- 15 March), restrictions (15 March -- 15 May), and
post-restriction (15 May -- end of September), and we focus our analysis on the states of California and Massachusetts. We adopt these
boundaries as a national-level reference, consistent with the period
during which the majority of U.S.\ states progressively enacted and
later eased shelter-in-place measures \cite{hale2021global}.

We measure the radius of gyration \cite{Gonzalez_08_rg} and compare distributions across different county-level political affiliations, degree of urbanisation, and pandemic phases. We fit power-law models to the tail of the trip-length distribution and compare the fitted tail parameters across the different phases. The rate of discovery of new places follows a diffusive power law \cite{scaling_visitation}, and the rank structure of visitation frequencies \cite{scaling_visitation, Schlapfer2021}. We further compute stop-point distributions in each trajectory's intrinsic reference frame \cite{Gonzalez_08_rg}. For trajectory complexity, we use three entropy measures --- random, Shannon, and Lempel-Ziv real entropy \cite{entropic_measures, Lu2013} --- which progressively add sensitivity to visit frequency, sequential structure, and the regularity of return behaviour \cite{Teixeira2021, Cuttone2016}. Whether complexity reduction during restrictions reflects a structural change in trajectories or simply fewer visited locations is one of the central questions we address. Finally, we apply a returner--explorer framework \cite{pappalardo_returners} to characterise shifts in habitual versus novel movement.

Our results show that the restriction phase was not a simple rescaling of pre-pandemic mobility. Individuals moved within a smaller spatial range, with a characteristic contraction scale of about 5~km. Short- and medium-range trip lengths were largely unchanged, and rare long-distance trips persisted. Mobility also reorganised around a small set of habitual locations. The discovery of new places slowed from a super-diffusive to a near-diffusive regime, and a larger share of visits concentrated on the two most frequented locations. Once scale and orientation were removed, the geometric shape of trajectories stayed essentially the same across phases. Trajectory entropy dropped during restrictions, but this drop is explained mainly by the smaller number of visited locations rather than by more regular or periodic routines. After restrictions were eased, mobility did not simply return to its pre-restriction state. The rebound was asymmetric, favoured shorter-range habitual mobility, and was more complete in California than in Massachusetts. Mobility began to decline before the statewide mandatory orders. Users in rural counties kept larger ranges than urban users throughout the period, while differences between county political groups were small. Taken together, these findings suggest that mobility-based epidemic models that only rescale baseline movement volumes or distances may miss relevant changes in exploration and visitation behaviour. Such features may need to be re-estimated when mobility data are used to describe behavioural regimes during large disruptions.

    \section*{Methods}
Following previous studies of mobility during the first wave of COVID-19~\cite{xiong2020mobile,chang2021mobility,lucchini} we partitioned the observation period into  three phases for descriptive convenience: pre-restriction (15 January–15 March), restriction (15 March–15 May), and later phases (15 May–30 September). These intervals do not imply uniform exposure to restrictions for the entire population, because the timing and stringency of policies varied across states and counties. The comparisons are observational and do not identify the causal effects of specific non-pharmaceutical restrictions. 
At the state level, the two states we study exhibit small departures
from this national timeline. California's first restriction was
locally targeted (six Bay Area counties, effective 17 March) before
becoming a statewide, mandatory order on 19 March, while Massachusetts issued a suggested, non-mandatory restriction on
23 March that never reached the mandatory stringency level observed in
California. As these offsets are small relative to the two-month span
of the restriction phase, and do not materially affect our results
(see Supplementary Section for a state-specific
sensitivity check for Massachusetts), we retain the national 15
March -- 15 May window as a common, uniformly-defined reference across
both states.

Generally speaking, these phases broadly correspond to i) the period immediately before the widespread implementation of non-pharmaceutical restrictions in the United States, ii) the interval during which mobility restrictions were broadly in place across most states, including California and Massachusetts, and iii) the subsequent phase of progressive restriction lifting and gradual reopening.

\subsection*{Mobility dataset}
\label{dataset}

We analysed privacy-enhanced, de-identified device-level GPS data from users that opted-in to share the data anonymously. The data is provided by Cuebiq Inc through the Cuebiq Data for Good COVID-19 Collaborative program, which provides access to de-identified and privacy-enhanced mobility data for academic research and humanitarian initiatives only. The provider reports that data collection and processing followed a CCPA and GDPR-compliant framework. To further preserve user privacy, the provider obfuscates home and work locations by assigning them to the centroid of the corresponding Census Block Group rather than reporting their exact coordinates. We focused our analysis on users residing in California and Massachusetts between 15 January and 30 September 2020. Our resulting preprocessed dataset contains trajectories from more than one million users in California and over $180,000$ users in Massachusetts. A summary of the statistics for each period are reported in Table~\ref{tabPeopleCal} for California, and Table~\ref{tabPeopleMA} in the Supplementary material for the case of Massachusetts. Since we focus on population-level statistical laws, we omit adjustments for phase-dependent mobile phone usage bias and evaluate distributions regardless of user overlap across the three periods. This introduces a limitation that we did not address: we are not considering the statistical change of individual behavior, but the statistical change of population behavior. 
Since user participation differs across phases, part of the observed population-level differences may reflect changes in sample composition or user activity in addition to changes in individual mobility behaviour.

Enhanced mobile phone data collected by Cuebiq is characterized by concentrated bursts of data collection within short time windows. The sampling process occasionally produced clusters of observations separated by short intervals, which could cause periods of intensive sampling to contribute disproportionately to trajectory statistics. We therefore applied a temporal thinning procedure (``time burst filter'') to the ordered stop-point sequence. Formally, for each trajectory $traj^{\alpha}$, two consecutive stop points satisfy,
\begin{equation}
    t_{\tau + 1} - t_{\tau} > t_{burst} 
\end{equation}
We define the time burst filter $t_{burst}$ as the minimum time interval within two successive stop points. Throughout this work we focused on the case of $t_{burst} = 1$ hour, and we also investigate the robustness of our results by considering thresholds of 8 and 24 hours, with the corresponding analyses, described in the Supplementary Material. 
The methods and the plots to reproduce these analysis are available on Github at \cite{amaduzzi2020humanmobilitycovid}

\subsection*{Notations}
\label{notations}

Table~\ref{tab:standard_definitions} summarizes the notation used throughout the manuscript. Here, we briefly introduce the main mathematical objects that define individual trajectories.
\begin{table}[H]
    \centering
    \begin{tabular}{>{\centering\arraybackslash}p{3cm} p{9cm}}
    \rowcolor{gray!30}
    \textbf{Symbol} & \textbf{Definition} \\
        \rowcolor{gray!5}
        $grid (i,j,k)$ & Geohash-7 grid and corresponding cell indices \\
    $d_g(\mathbf{x},\mathbf{y})$ & Geodesic distance between locations $\mathbf{x}$ and $\mathbf{y}$\\
        \rowcolor{gray!5}
        $Users$ & Set of users\\
        $Trajs$ & Set of all trajectories\\
        \rowcolor{gray!5}
        $\alpha,\beta$ & User indices\\
        $l^\alpha$ & Length of the trajectory of user $\alpha$ (km)\\
        \rowcolor{gray!5}
        $T^\alpha$ & Set of timestamps associated with the stop points of trajectory $\alpha$\\
        $|T^\alpha|$ & Duration of trajectory $\alpha$ (hour)\\
        \rowcolor{gray!5}
        $T_{\mathrm{start}},\,T_{\mathrm{end}}$ & Beginning and end of the observation period\\
        $T^{p}_{\mathrm{start}},\,T^{p}_{\mathrm{end}}$ & Beginning and end of pandemic phase $p$\\
        \rowcolor{gray!5}
        $t^{\alpha}_{\mathrm{start}},\,t^{\alpha}_{\mathrm{end}}$ & Start and end time of trajectory $\alpha$ \\
        sp & Stop point \\
        $\tau$ & Temporal index of a stop point\\
        \rowcolor{gray!5}
        $t_{sp}$ & Minimum stop duration (10 min)\\
        $\delta t_t$ & Residence time at stop point $sp_t$ (hour)\\
        \rowcolor{gray!5}
        $Users^p$ & Set of users active during period $p$\\
        $Trajs^p$ & Set of trajectories during period $p$\\
    \end{tabular}
    \caption{Notation and definitions.}
    \label{tab:standard_definitions}
\end{table}
Each user $\alpha$ is associated with a trajectory
\begin{equation}
    traj^{\alpha} \;=\; \{sp^{\alpha}_t\}_{t \in T^{\alpha}}
\end{equation}
where each stop point,
\begin{equation}
    sp^{\alpha}_t \;=\; (lat^{\alpha}_t,lon^{\alpha}_t,t,\hat{t})
\end{equation}
Represent a location visited continuously for at least $t_{sp}$, with starting time $t$ and ending time ($\hat{t}$) such that the signal of the mobile phone remains for a minimum duration of $10$ minutes. We consider the time of the stop point as the beginning of the stop.  The residence time at each stop is,
\begin{equation}
    \delta_t \;=\; \hat{t} - t
\end{equation}

Since we divide the overall period into three different periods, p,  we represent the different stages of the stay-at-home restrictions with the indexes $\{pre, \; during, \; post\}$ and generically referred to the periods with $p$. For each user $\alpha$, we constructed a separate sub-trajectory $traj^{\alpha,p}$ from stops whose start times fell within period $p$. Stops spanning multiple periods were excluded from the analysis. A user, $Users^{p}$, was included in each of the different phases periods ,$p$, if the resulting trajectory satisfied the inclusion criteria within each of the partitions.

The geodesic distance is defined as $d_g(\vec{x},\vec{y})$ between two points $\vec{x}$ and $\vec{y}$. We define the mobility network of user $\alpha$ as $G^{\alpha} = (N^{\alpha},E^{\alpha})$.
This results in a weighted directed graph where the nodes $N$ are the stop points and the edges $E$ are the transitions between successive stop points with the weight of edges $w_{e}$ defined as the number of time in which the transition happened. Notice that the graph is naturally loop-less since successive points in the same location are considered as an extended stay of longer $\delta_t$. We call $k^{\alpha,in}_n$ the number of time node $N$ is visited.

\subsection*{Mobility metrics}

To characterize the changes in human mobility during the three phases, we compute complementary metrics that measure: (i) the spatial extent of mobility, (ii) the exploration and visitation behaviour of individuals, and (iii) the complexity of individual trajectories.

We begin by quantifying the spatial extent of individual mobility through a modified duration-weighted version of the radius of gyration~\cite{centellegher2025job,kalleitner2022varieties}, shown in Eq.~\ref{rg}. This definition weights each stop by the time spent at that location, thus longer stays contribute proportionally more to the centre of mass and spatial dispersion. The time-weighted centre of mass $\vec{x}^{\alpha}_{cm}$ is defined as
\begin{equation}
    \vec{x}^{\alpha}_{cm} = \frac{\sum_{t \in T^{\alpha}} \delta t_{t}\, \vec{x}^{\alpha}_{t}}{\sum_{t \in T^{\alpha}} \delta t_{t}},
    \label{cm}
\end{equation}
and the radius of gyration is then,
\begin{equation}
     R^{\alpha}_{g} = \sqrt{\frac{\displaystyle\sum_{t \in T^{\alpha}} \delta t_{t}\; d_{g}(\vec{x}^{\alpha}_{t},\vec{x}^{\alpha}_{cm})^2}{\displaystyle\sum_{t \in T^{\alpha}} \delta t_{t}}}
     \label{rg}
\end{equation}
where $\delta t_{t}$ is the time spent at the stop point of coordinates $\vec{x^{\alpha}_{t}}$, $\vec{x^{\alpha}_{cm}}$ is the time-weighted centre of mass of the full trajectory (Eq.~\ref{cm}), and $d_{g}$ is the geodesic distance projected onto the local tangent plane. All distances are expressed in kilometres. The centre of mass is computed once over the entire observation period for each user--period pair; it is not updated incrementally as new stops are observed. 
The choice of a duration-weighted radius of gyration is motivated by epidemiological considerations: we are not only interested in how far people travel, but also in how long they remain at the places they visit, since dwell time is itself relevant to the probability of disease transmission~\cite{ferretti2024digital,lucchini2026implications}.

Now, to characterise the geometry of individuals trajectory independently of their orientation and spatial extent, we analyse the distribution of stop points following the work of Gonzalez et al.~\cite{Gonzalez_08_rg}. Each trajectory is projected onto a local tangent plane, centred on its centroid, rotated along its principal axes, and normalized by the corresponding standard deviations. 
Let ($x_{\alpha,t}$,$y_{\alpha,t}$) denote the projected Cartesian coordinates of the stop points. After centring the trajectory around its centroid ($\tilde{x}_t = x_t^{\alpha} - \bar{x}^{\alpha}$, $\tilde{y}_t = y_t^{\alpha} - \bar{y}^{\alpha}$), we compute the inertia tensor,
\begin{align}
    I_{xx} = \sum_t \tilde{x}_t^2,\quad,
    I_{yy} = \sum_t \tilde{y}_t^2,\quad,
    I_{xy} = \sum_t \tilde{x}_t\tilde{y}_t.
\end{align}
We find the principal rotation angle $\theta$ that diagonalises the tensor
\begin{equation}
    \cos\theta =
    \frac{-I_{xy}}
    {\left(\frac{I_{xx}-I_{yy}+\mu}{2}\right)
    \sqrt{1+\frac{I_{xy}^2}
    {\left(\frac{I_{xx}-I_{yy}+\mu}{2}\right)^2}}}
\end{equation}
where $\mu$ is defined as:
\begin{equation}
    \mu = \sqrt{4I_{xy}^2 + (I_{xx} - I_{yy})^2}
\end{equation}
Then, we rotate to the principal frame: $(x'_t, y'_t) = R_\theta\,(\tilde{x}_t, \tilde{y}_t)$,
and compute the root-mean-square spreads: $\sigma_{x'} = \sqrt{\langle x'^2\rangle}$, $\sigma_{y'} = \sqrt{\langle y'^2\rangle}$, to finally normalised $\hat{x}_t = x'_t/\sigma_{x'}$, $\hat{y}_t = y'_t/\sigma_{y'}$.

The resulting two-dimensional probability distribution of the normalized coordinates ($\hat{x},\hat{y}$) provides a representation of trajectory geometry that is standardized with respect to overall scale and principal-axis orientation, enabling direct comparison across users and pandemic periods.

To quantify exploratory behaviour, we analyse both the temporal accumulation of newly visited locations $S(t)$ and the visitation frequency of the most frequently visited places by individuals. Let $r$ denote the rank $r = 1,\ldots,r_{max} = 10$.
To compare users with a significantly different number of visited locations, we limited our analysis to the $r_{max}$ most visited places in the period considered for those who had more then $r_{max}$.
The normalised visit frequency at rank $r$ in period $p$ is given by Eq.~\ref{k_avg}.

\begin{equation}
    \langle k \rangle^{r,p} =  \left\langle \frac{k^{\alpha,r}}{\sum_{r' = 1}^{\min(10,\,r_{\max})}k^{\alpha,r'}} \right\rangle_{{\alpha} \in {Trajs^p}}
    \label{k_avg}
\end{equation}
where, $p$ denotes the time period, $r$ represents the rank of a node in an individual’s mobility network (with $r=1$ being the most visited location), and $k^{\alpha, r}$ is the number of trips made by individual $\alpha$ to the node of rank $r$. The distribution is computed separately for each period $p$, considering only the individuals active during that period. This metric captures how the frequency of visits to the most important locations changes across the population over different time periods.

We additionally characterize the location diversity, visitation-frequency heterogeneity, and sequential regularity of mobility trajectories using three complementary entropy measures.

\begin{equation}
    H_{\text{random}} = \log_2 n \label{rdm_entropy}
\end{equation}
\begin{equation}
    H_{\text{uncorrelated}} = -\sum_{i=1}^n p_i \log_2 p_i
    \label{unc_entropy}
\end{equation}
\begin{equation}
    H_{\text{real}} = \left(\frac{1}{T}\sum_{t=1}^{T} \Lambda_t\right)^{-1} \log_2 T
    \label{real_entropy}
\end{equation}
The random entropy (Eq.~\ref{rdm_entropy}) depends only on the number $n$ of distinct visited locations. The uncorrelated entropy (Eq.~\ref{unc_entropy}) adds sensitivity to the relative frequency of visits $p_i$, treating successive stops as independent. The real entropy (Eq.~\ref{real_entropy}) is a Lempel-Ziv estimator that accounts for the sequential order of visits and converges to the true entropy rate of the trajectory \cite{Lempel-Ziv, entropic_measures}. Here $T$ is the total number of stop points and $\Lambda_t$ is the length of the shortest substring starting at position $t$ that has not previously appeared in positions $1, \ldots, t-1$. These three measures provide a progressively finer view of trajectory complexity. 

Finally, let $V_{\alpha,p}^{k}$ be the set of the $k$ most frequently visited locations. The $k$-radius of gyration was calculated using only stops assigned to those locations: the $k$-radius of gyration extends Eq.~\ref{rg} by restricting the computation to $V_{\alpha,p}^{k}$, quantifying the spatial dispersion only of stops occurring among the $k$ most frequently visited locations following \cite{pappalardo_returners}. 
\subsection*{The California case study}

Here we summarize in Table~\ref{tabPeopleCal} the statistics for the case of California in the different periods.
{
\renewcommand{\arraystretch}{1.5} %
\setlength{\tabcolsep}{12pt}      %
\begin{table}[H]
    \centering
    \begin{tabular}{l l r r r}
        \rowcolor{gray!5}
        Period & Dates & $N_c$ & $T_c$ & $\langle T_c \rangle_p$ \\
        \rowcolor{figblue!20}
        Pre- restrictions & 15 January -- 15 March & 1,009,464 & 164,994,721 & 163 \\
        \rowcolor{figorange!20}
        During- restrictions & 15 March -- 15 May     &   490,673 &  12,701,431 & 26 \\
        \rowcolor{figgreen!20}
        Post- restrictions & 15 May -- 30 September    &   460,654 &  32,096,311 & 70 \\
    \end{tabular}
    \caption{California preprocessed dataset with the columns of the number of users $N_c$, the number of total points $T_c$ and the average points per each trajectory $\langle T_c \rangle_p$.}
    \label{tabPeopleCal}
\end{table}
}
        
    \section*{Results}
        \label{results}

In this section, we characterise differences in observed mobility patterns across three calendar windows in 2020, comparing pre-restriction phase, restriction-phase, and first-reopening phase windows. As mentioned previously, our analysis covers three different aspects: i) the spatial extent of mobility, ii) changes in visitation and exploration patterns, and iii) the predictability of individual trajectories. As a last step, we also explore whether these behavioural changes differed across population groups, by considering the county-level degree of urbanisation and political classification of users counties or residence. Here, we focus our analysis on the case of California, while results for the Massachusetts metropolitan area are reported in the Supplementary Material. Across all the metrics considered, the qualitative behaviour is consistent between California and Massachusetts (see Supplementary Material), with only slight differences between the two cases. 
The Supplementary Material also includes a stability analysis for both California and Massachusetts, testing the robustness of our results to the choice of the burst-filter threshold $t^{burst} \in \{1, 8, 24\}$~h (see the Sensitivity Analysis Section) and showing that qualitative features are robust across different values of $t_{burst}$.

\subsection*{Contraction of mobility during social-distance restrictions}

\begin{figure}[h!]
    \includegraphics[width=1\textwidth]{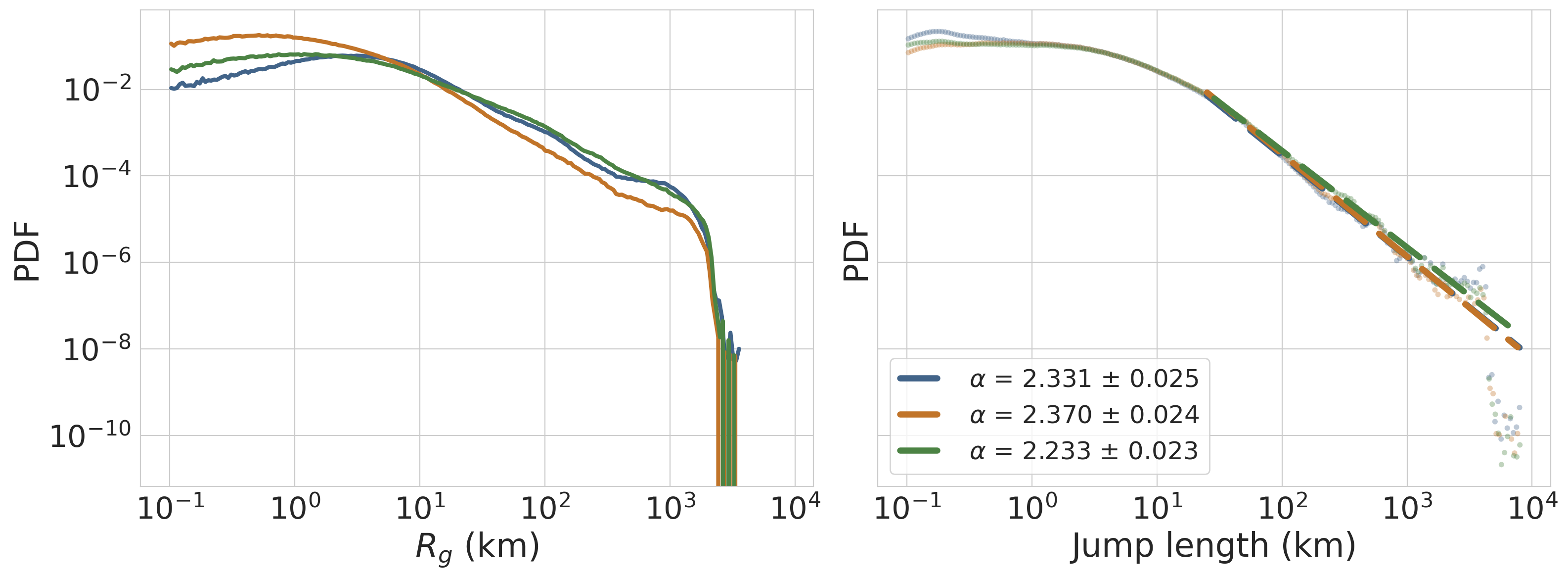}
    \caption{{\bf Changes in mobility across the pandemic phases of social distancing}. We consider three different phases according to the social-distances restrictions implemented in the early stages of the pandemic. Lines in blue correspond to the period before the restrictions (from 15 of January to 15 of March), lines in orange correspond to the period during distancing restrictions (from 15 of March to 15 of May), and lines in green correspond to the restriction period after the restrictions (from 15 of may to September). The time threshold consider for the trajectories is one hour. Plot on the left corresponds to the {\it distribution of radius of gyration}, where we can see that the restriction period is characterised by a shift from medium to short-distance mobility, while long-distance trips remain largely unchanged. Plot on the right corresponds to the {\it distribution of jump lengths} between successive stop points, with power-law fits to the tail yielding exponents $\alpha_{\text{pre}} = 2.331 \pm 0.025$, $\alpha_{\text{during}} = 2.370 \pm 0.024$, and $\alpha_{\text{post}} = 2.233 \pm 0.023$; the three distributions overlap closely below $\sim$10--20~km, with differences emerging only in the tail.}
    \label{radg_and_dist_california}
\end{figure}

We began our analysis by examining, in Fig.~\ref{radg_and_dist_california}, the distribution of the duration-weighted radius of gyration (Eq.~\ref{rg}, left) and the distribution of trip distances (right). 
When comparing the three restriction periods, we observe that the distribution of the radius of gyration shifts toward smaller values during restrictions (orange, compared with the blue and green curves). Because each stop is weighted by its duration, this shift indicates that a larger proportion of observed dwell time was distributed within a smaller spatial extent around each $User_p$ trajectory’s centre of mass: individuals tended to move within a smaller geographical area and to remain there for less time. This effect is present in both datasets and is more pronounced in the case of Massachusetts (Fig. \ref{radg_and_dist}). The scale at which this over-representation emerges is $\sim$5~km. We are not able to  distinguish whether this concentration resulted from changes in work, schooling, trip purpose, destination availability, perceived risk, or other factors since the available data does not contain these informations.

The distribution of single-trip distances shows a largely invariant pattern across the three periods: below $\sim$10--20~km, the pre-, during-, and post-restriction curves overlap closely, indicating that the shape of short- and medium-range jump lengths is not substantially altered by the restrictions. Differences between periods emerge mostly in the long-distance tail ($\ge 20 km$). The fitted power-law exponents in this tail are $\alpha_{\text{pre}} = 2.331 \pm 0.025$, $\alpha_{\text{during}} = 2.370 \pm 0.024$, and $\alpha_{\text{post}} = 2.233 \pm 0.023$: the during-restriction exponent is the highest of the three, indicating a slightly lighter tail, i.e.\ long trips are marginally rarer relative to the bulk of the distribution, during restrictions, while the post-restriction exponent is the lowest, indicating the heaviest tail among the three periods. These differences are modest compared to the uncertainties reported, and the overall persistence of rare long-distance trips across all periods is the more robust qualitative finding. 
Changing the time filter to 8 and 24 hours (Table~\ref{tab:distance_powerlaw}), we notice that the power-law exponent is lower than in the 1-hour case across all three periods. This indicates that, at this coarser temporal resolution, the tail of the trip-distance distribution appears heavier, i.e.\ long trips become relatively less rare, likely because the wider burst filter merges together some of the short, closely-spaced stops that at the 1-hour threshold would otherwise be resolved as separate, shorter displacements --- effectively reclassifying part of the medium-range mobility as long-range jumps \cite{gallotti2018tracking}. Despite this shift in absolute exponent values, the ordering across periods is preserved: the during-restriction exponent remains the highest and the post-restriction exponent the lowest at both thresholds, indicating that the relative pattern across pandemic phases is robust to the choice of $t^{burst}$, even though the absolute tail weight is not. %

Overall, the restriction-phase was characterized by smaller duration-weighted radii of gyration and a modestly lighter long-distance tail. Short- and medium-range jump-length distributions remained broadly similar, and rare long-distance movements persisted. 
The power law fit is compatible with the observations made in \cite{ferres2026powerlawhumanmobility}. 
However, this does not fully explain how individuals mobility behaviour changed. Shorter trips may occur more often because of a relative reduction in the number of trips or because individuals changed their routine concentrating their activity around a smaller set of familiar locations. To better understand this, we further investigate the relationship between the radius of gyration and the k-radius of gyration, together with visitation and exploration patterns.

\subsection*{Reduced exploration and increased concentration around habitual locations}

To better understand the behavioural changes underlying the reduction of mobility, here we investigate how individuals distributed their trips across locations. 

\begin{figure}[h!]
    \centering
    \includegraphics[width=1\textwidth]{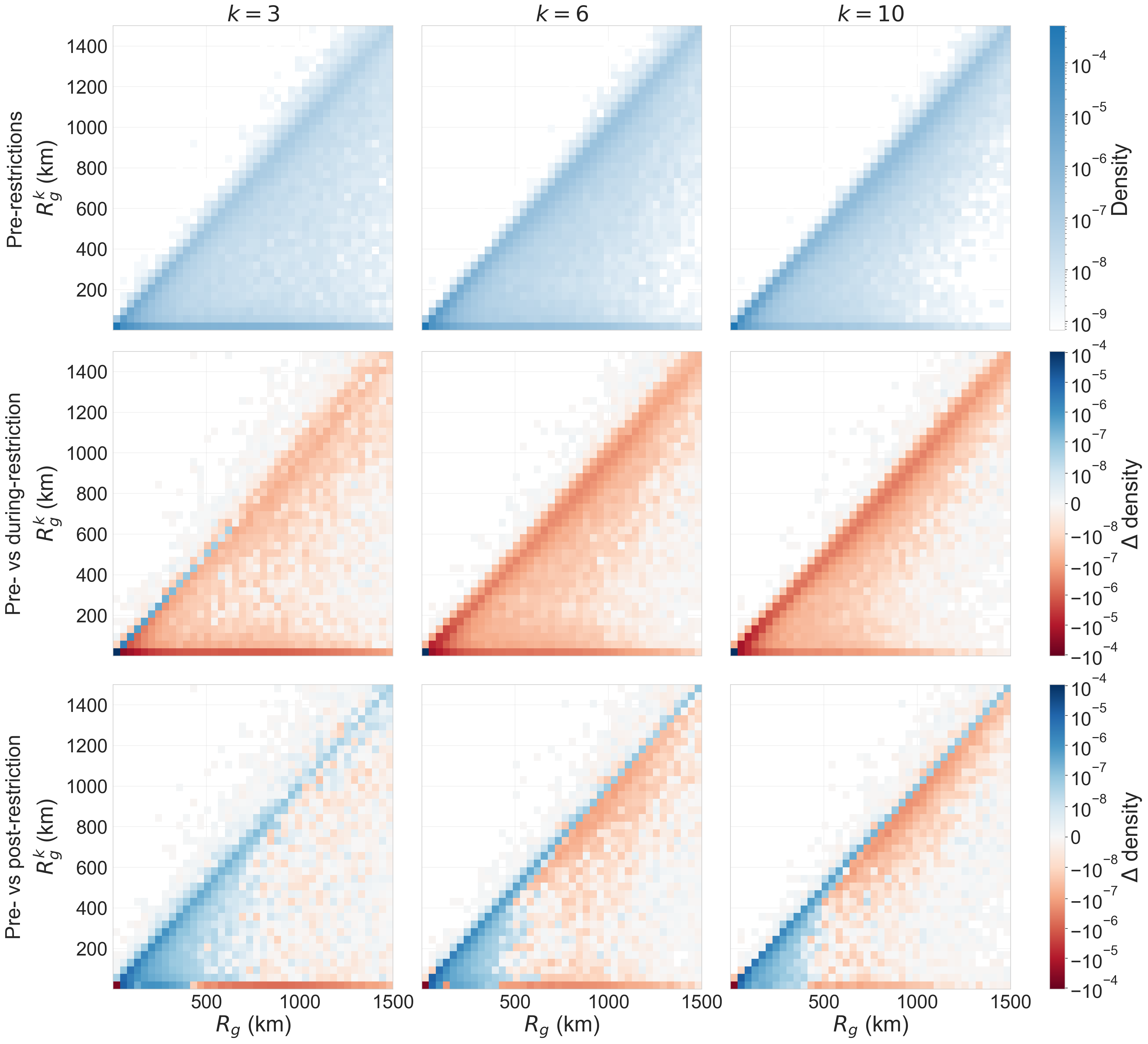}  
    \caption{{\bf Change in the joint $R_g$--$R_g^{k}$ distribution relative to the pre-restriction baseline.} The top row reproduces the pre-restriction (15~Jan--15~Mar) joint density as a reference. The middle and bottom rows show, on the identical $(R_g, R_g^{k})$ bin grid, the absolute density difference between the during-restriction and post-restriction periods and that same baseline, respectively. Panels show $k = 3, 6, 10$ (left to right). Blue indicates a gain in density relative to the baseline, red a loss. Values close to $R_g^{k} = R_g$ indicate that the $k$ most visited locations recover most of the trajectory's spatial dispersion, whereas $R_g^{k} \ll R_g$ indicates that less frequently visited locations contribute substantially to the full radius of gyration. 
    During restrictions, probability shifts strongly toward small $R_g$ and $R_g^{k}$, indicating an overall spatial contraction, while configurations with large $R_g$ but small $R_g^{k}$ become less common, consistent with the suppression of large excursions outside frequently visited locations.
    Post-restriction, the pattern is not a simple reversion to baseline: density along and near the diagonal $R_g^{k} \approx R_g$ increases at small-to-moderate $R_g$ (blue), while the largest-$R_g$ region remains suppressed relative to baseline (red), indicating a rebound concentrated in shorter-range, habitual mobility, with long-distance travel still lagging behind pre-restriction levels.}            
    \label{returner_explorers_california}
\end{figure}

In Fig.~\ref{returner_explorers_california}, we examine how the radius of gyration relates to the $k$-radius of gyration $R_g^{k}$, i.e.\ how much of a user's full spatial dispersion is already recovered when only their $k$ most frequently visited locations are considered. Rather than compare the three periods' raw joint densities directly, we express the during- and post-restriction densities as differences from the pre-restriction baseline on the identical $(R_g, R_g^{k})$ grid (middle and bottom rows), alongside the baseline itself (top row).

\begin{figure}[h!]
    \centering
    \includegraphics[width=1\textwidth]{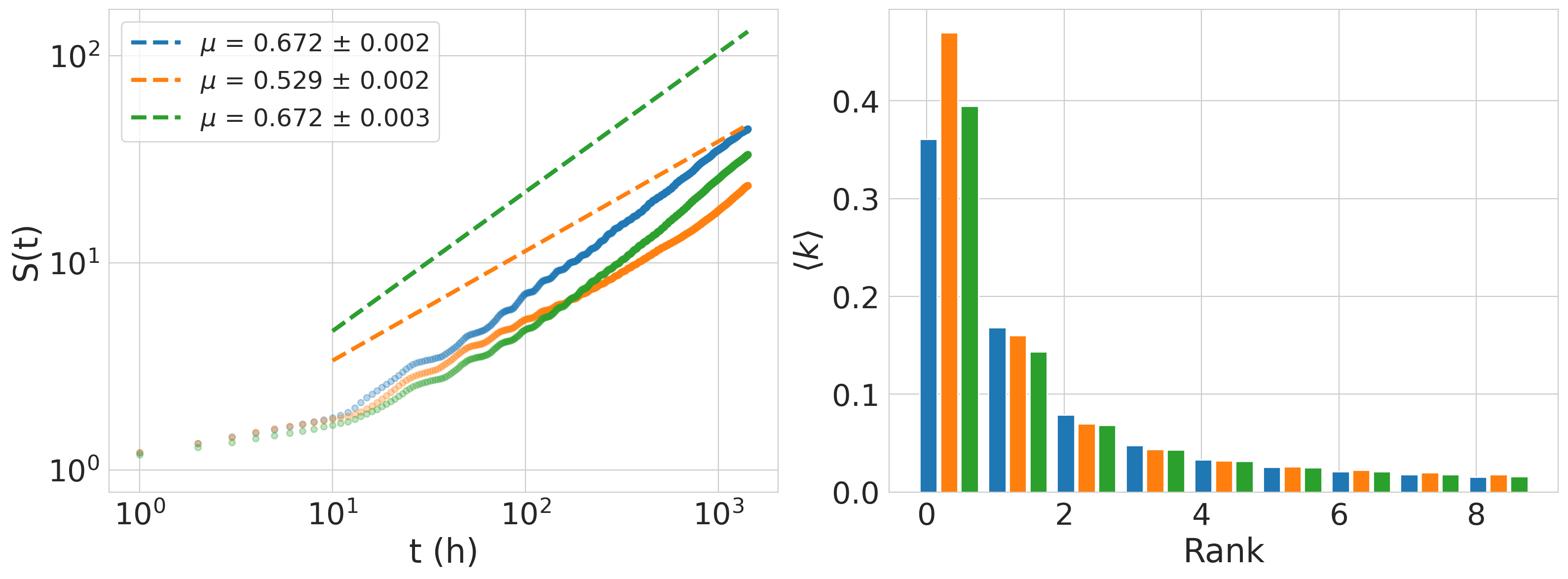}
    \caption{{\bf Exploration rate and visitation-frequency ranking.} Colors correspond to the pre-, during-, and post-restriction periods, as in Fig.~\ref{radg_and_dist_california}. Left: average number of distinct stop points discovered over time, $S(t)$, with power-law fits $S(t) \sim t^{\mu}$ yielding $\mu_{\text{pre}} = 0.672 \pm 0.002$, $\mu_{\text{during}} = 0.529 \pm 0.002$, and $\mu_{\text{post}} = 0.672 \pm 0.003$; discovery during restriction is markedly closer to diffusive ($\mu = 0.5$) than the super-diffusive pre- and post-restriction regimes. Note that the pre and post fit overlap. Right: normalised visit frequency by location rank for the ten most-visited places per user.}
    \label{visitation_increase-frequency_california}
\end{figure}

During restriction, density shifts markedly toward $R_g^{k} \approx 0$ already at $k=3$: the share of users whose top-$k$ locations alone place them in the smallest $(R_g, R_g^{k})$ bin rises from $73\%$ pre-restriction to $88\%$ during restriction ($92\%$ in Massachusetts), at the expense of density everywhere else in the plane. 
Overall, restrictions suppress the spatial dispersion generated outside the set of frequently visited locations, together with an overall contraction of mobility. Post-restriction, this share falls to $65\%$ ($71\%$ in Massachusetts) -- below its pre-restriction level -- but the recovered density is not evenly distributed: it concentrates along the diagonal $R_g^{k} \approx R_g$ at small-to-moderate $R_g$, while the largest-$R_g$ region remains below baseline, pointing to an asymmetric rebound weighted toward shorter-range, habitual mobility rather than a uniform return to pre-restriction patterns. By looking at the discovery pattern in Fig.~\ref{visitation_increase-frequency_california}, we observe that people tend to discover fewer places during restriction, becoming more returners than explorers~\cite{pappalardo_returners}. 
This is confirmed quantitatively by the exponent of the newly visited places (discovery curve) $S(t) \sim t^{\mu}$, which drops from $\mu = 0.672 \pm 0.002$ (pre-restriction, super-diffusive) to $\mu = 0.529 \pm 0.002$ (during restriction, close to diffusive) and returns to $\mu = 0.672 \pm 0.003$ (post-restriction). The frequency of visitation also concentrates more heavily in the top-ranked locations during restriction: the two most-visited locations account for 62\% of visits during restrictions, against 53\% in both the pre- and post-restriction periods, while the remaining ranks are relatively unaffected. Together with Fig.~\ref{returner_explorers_california}, this suggests that during restriction trajectories are organised around a small number of habitual locations acting as attractors, an effect that is more pronounced in Massachusetts, where the single most-visited location alone accounts for nearly half of all visits during restrictions (Fig.~\ref{visitation_increase-frequency_massachusetts}).

\begin{figure}[h!]
    \centering
    \begin{subfigure}[b]{\textwidth}
        \centering
        \includegraphics[width=0.9\textwidth]{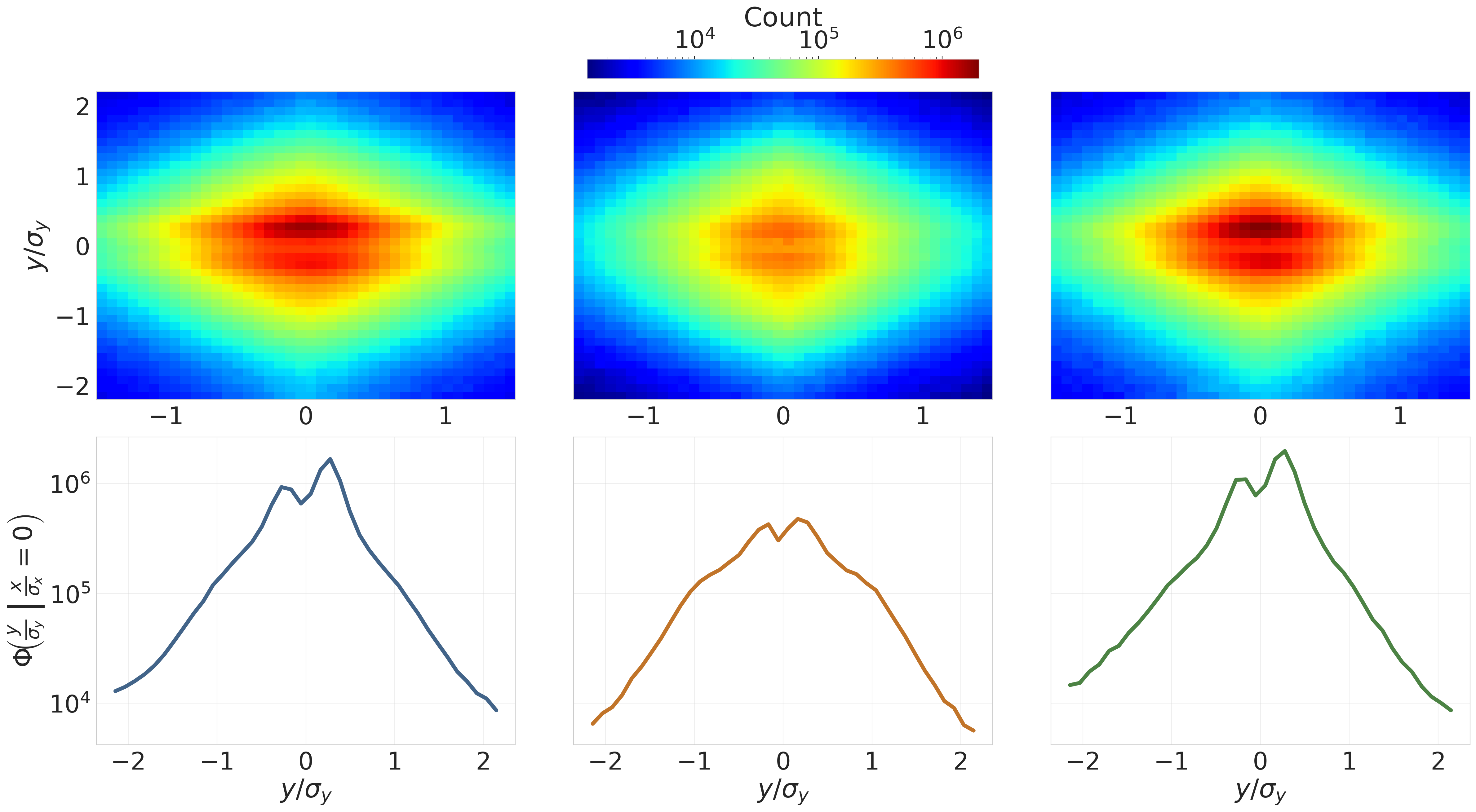}
        \phantomcaption
        \label{Gonzalez_california}
    \end{subfigure}
    \vskip 1em
    \begin{subfigure}[b]{\textwidth}
        \centering
        \includegraphics[width=0.9\textwidth]{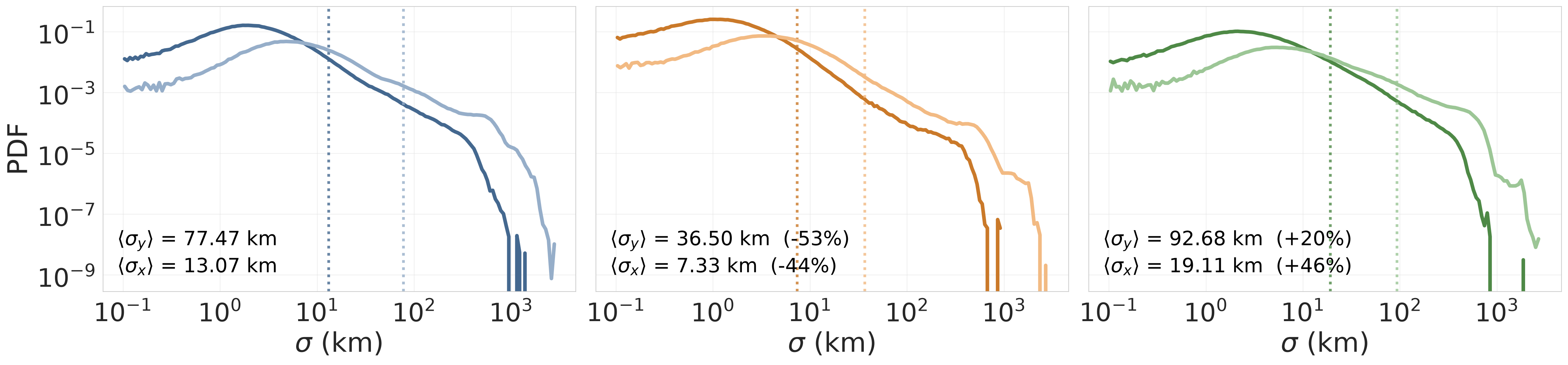}
        \phantomcaption
        \label{sigma}
    \end{subfigure}
    \caption{{\bf Trajectory shape and spread along the intrinsic reference frame.} Panels correspond, left to right, to the pre-, during-, and post-restriction periods within each subfigure, as in Fig.~\ref{radg_and_dist_california}. (Top) Top row: two-dimensional distribution of stop-point positions in each trajectory's principal-axis frame, rescaled by the standard deviations of the rotated coordinates. Bottom row: conditional probability density along $\hat{y}$ at $\hat{x}=0$. The overall elongated shape is preserved across periods, with the density concentrated closer to the origin during restriction and re-spreading in the post-restriction period, consistent with the changes in $\sigma_{x'}$, $\sigma_{y'}$ reported below. (Bottom) Distribution across users of the standard deviations $\sigma_{x'}$ (magenta) and $\sigma_{y'}$ (cyan) of stop-point positions along each trajectory's principal axes, in kilometres. Both mean values drop substantially during restriction relative to the pre-restriction period ($\langle\sigma_{x'}\rangle$: $13.12 \to 7.29$~km, $-44\%$; $\langle\sigma_{y'}\rangle$: $77.64 \to 36.49$~km, $-53\%$). In the post-restriction period, both axes recover beyond their pre-restriction values ($\langle\sigma_{x'}\rangle = 19.18$~km, $+46\%$; $\langle\sigma_{y'}\rangle = 93.01$~km, $+20\%$, relative to the pre-restriction period). Both post-period means exceed their corresponding pre-period means.}
    \label{traj_shape_and_spread}
\end{figure}

These trips are distributed in 2-dimensional space in an invariant shape (Fig.~\ref{Gonzalez_california}), but with density peaks closer to the origin during restrictions, corresponding to a $44$--$53\%$ reduction in the spread along the two principal axes (Fig.~\ref{sigma}). In the post-restriction period, both axes not only recover but exceed their pre-restriction spread by $20$--$46\%$,
California showed a higher mean spread along both axes during the third phase than in the pre-restriction period, whereas Massachusetts (Fig.~\ref{sigma_massachusetts}) showed recovery along one axis and a persisting reduction along the other.
\subsection*{Reduced location diversity was associated with lower trajectory entropy}

\begin{figure}[h!]
    \centering
    \includegraphics[width=1\textwidth]{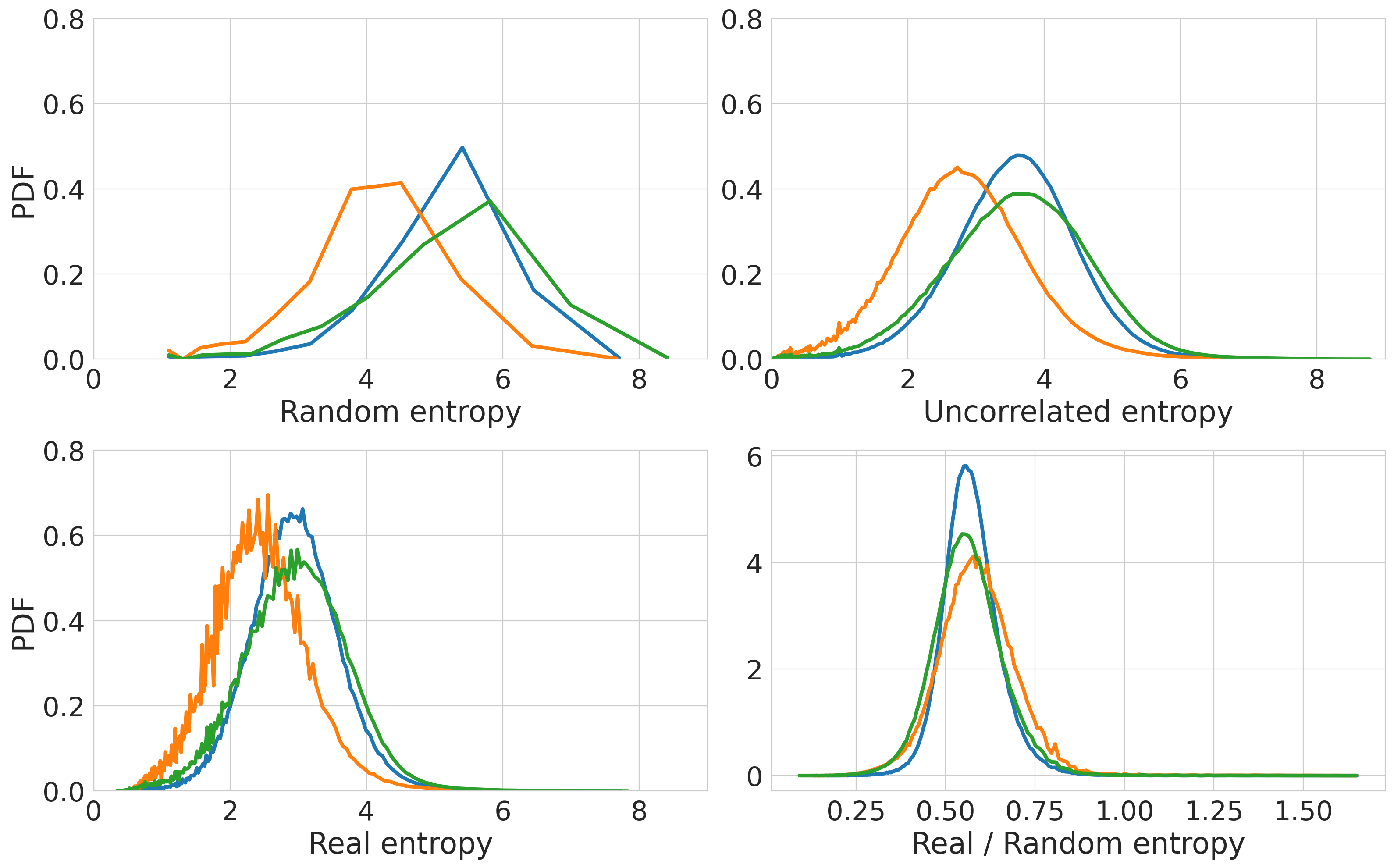}
    \caption{{\bf Entropy measures of individual trajectories.} Colors correspond to the pre-, during-, and post-restriction periods, as in Fig.~\ref{radg_and_dist_california}. Distributions across users of random entropy $H_{\text{random}}$ (top left), uncorrelated entropy $H_{\text{uncorrelated}}$ (top right), and real entropy $H_{\text{real}}$ (bottom left); all three shift toward lower values during restriction, indicating reduced trajectory complexity. Bottom right: distribution of the per-user ratio $H_{\text{real}}/H_{\text{random}}$, which shifts toward higher values (closer to 1) during restriction relative to the pre- and post-restriction periods, indicating that the reduction in complexity is driven mainly by the smaller number of visited locations rather than by increased sequential structure. Corresponding Kolmogorov-Smirnov statistics are reported in Table~\ref{tab:ks_entropy}.}
    \label{entropic_measures_california}
\end{figure}

The reduced amount of information conveyed by trajectories during restrictions is also visible from entropic considerations (Fig.~\ref{entropic_measures_california}). Across all three entropy measures, we find that trajectories exhibit lower entropy during restrictions than before or after, indicating that mobility patterns become less diverse. This reduction can be explained by two, non-exclusive, mechanisms. First, individuals may simply visit fewer distinct locations, which naturally reduces entropy by limiting the number of possible places that can be visited. Second, mobility routines may become more regular, with people visiting the same locations in a more repetitive order, producing lower estimated entropy through greater regularity in the observed location sequence.

To better understand the origin of this reduction, we examine the three entropy measures separately. The random entropy (Eq.~\ref{rdm_entropy}) depends only on the number of distinct locations $n$ visited by a user; its drop during restrictions is a direct signature of the smaller $n$ observed in this period (Fig.~\ref{returner_explorers_california}). The uncorrelated entropy (Eq.~\ref{unc_entropy}) additionally accounts for how visits are distributed across those $n$ locations; since visitation concentrates more heavily in the top-ranked places during restrictions (the two most-visited locations account for $62\%$ of visits, against $53\%$ pre- and post-restriction, Fig.~\ref{visitation_increase-frequency_california}), this skew lowers it further, beyond what the drop in $n$ alone explains. The real entropy (Eq.~\ref{real_entropy}), estimated via the Lempel-Ziv algorithm, additionally captures sequential order; if restrictions trajectories were becoming intrinsically more periodic, the real entropy should fall by \emph{more} than a simple drop in $n$ would produce.

To test this last point directly, we compute the per-user ratio between the real and random entropy (Eq.~\ref{rdm_entropy}, \ref{real_entropy}), shown in Fig.~\ref{entropic_measures_california} (bottom right). Two limiting cases clarify how to read this ratio:
\begin{itemize}
    \item \textbf{No exploitable sequential structure:} If the sequence of visited locations were statistically indistinguishable from i.i.d.\ uniform draws over the $n$ visited locations, the Lempel-Ziv estimator of the real entropy converges to the random entropy itself. In this limit, the real-to-random ratio tends to $1$: a ceiling corresponding to no order beyond what is already fixed by $n$.

    \item \textbf{Fully periodic structure:} If instead the sequence followed a short deterministic cycle (e.g.\ a period-2 home$\leftrightarrow$work pattern), the Lempel-Ziv parsing compresses almost perfectly, and the real entropy tends to $0$. In this limit, the real-to-random ratio tends to $0$: a floor corresponding to maximal exploitable sequential structure.
\end{itemize}
Given these two reference points, we can interpret the observed shift: the distribution of the real-to-random entropy ratio across users moves \emph{closer to 1} during restrictions, relative to the pre- and post-restriction periods. Since a ratio of $1$ is the no-additional-structure ceiling and $0$ is the maximal-structure floor, a shift toward $1$ means that users are, on average, moving closer to the ceiling, i.e.\ \emph{less} additional order is being exploited beyond what $n$ alone fixes. Equivalently, the gap between the random and real entropy,  the portion of entropy specifically attributable to sequential order, shrinks during restrictions rather than growing.

Combined with the drop in random entropy itself (fewer places visited), the full picture is that the absolute drop in real entropy observed in Fig.~\ref{entropic_measures_california} is driven almost entirely by the drop in $n$: the residual structural, sequential predictability does not increase during restrictions, and if anything decreases slightly relative to what the smaller $n$ alone would predict. We therefore conclude that the reduction in information content observed across all three measures is explained primarily by the smaller number of distinct places visited, rather than by trajectories becoming intrinsically more periodic or structured in their sequence of visits. 

\subsection*{Rural and political heterogeneity in mobility dynamics}

In this section, we stratify the population according to two characteristics: the rurality level of the area of residence and the governing political party of the county. Rural areas are defined as those with a population density below $500$ people per $km^2$.
We split users having at least two visited places in the considered week indipendently and compute the radius of gyration of the cut mobility for that week, so that different weeks contain independent mobility.
\begin{figure}[h!]
    \centering
    \includegraphics[width=1\textwidth]{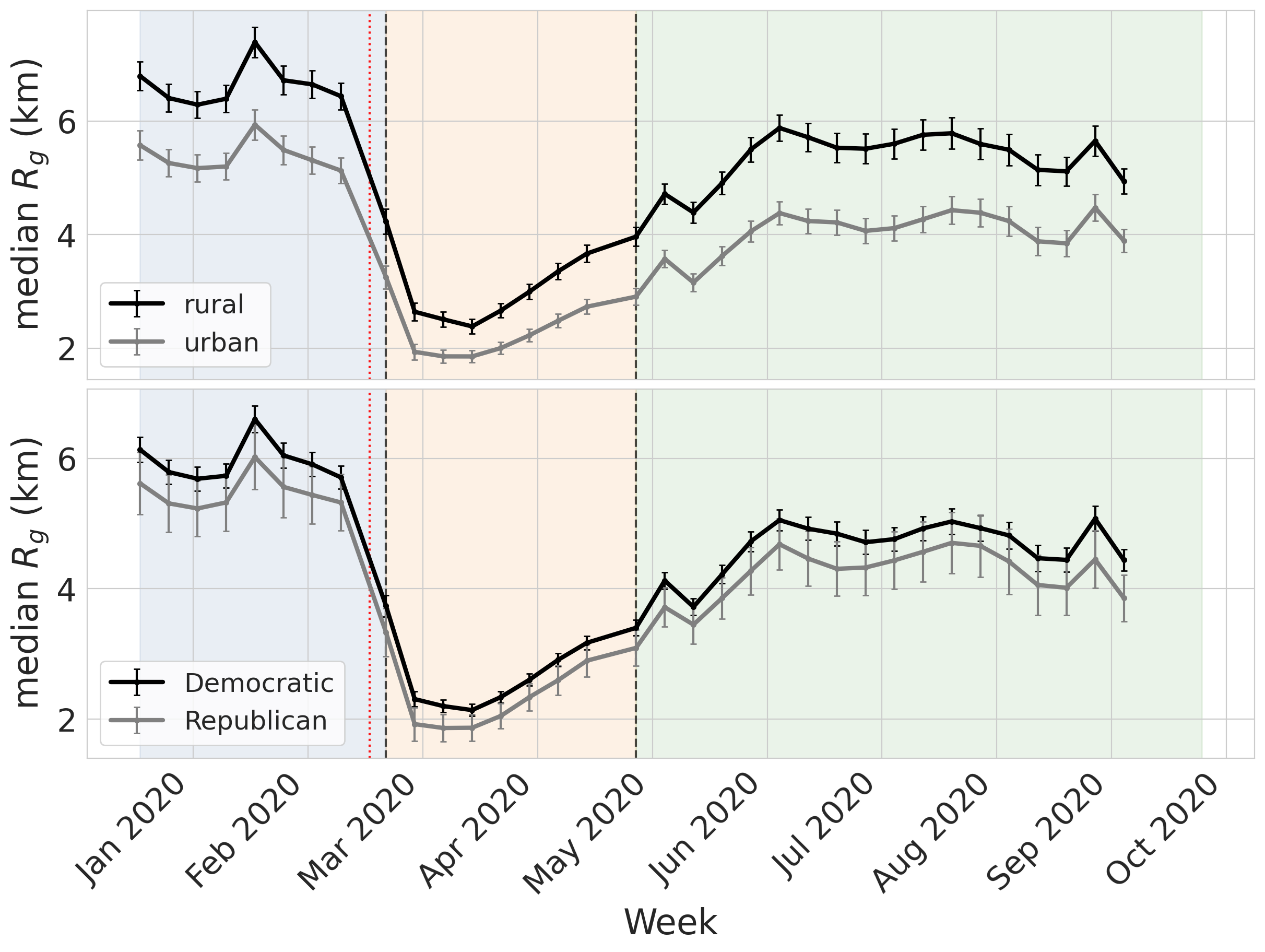} 
    \caption{{\bf Weekly radius of gyration by political affiliation and rurality}. Vertical lines mark the boundaries of the pre-, during-, and post-restriction periods. In both panels, error bars represent the standard error of the mean, and the red vertical line marks March 11 , when non-mandatory restrictions were recommended. Top: average weekly radius of gyration by governing party of the home county; party affiliation has little effect on the depth of mobility reduction during restrictions but is associated with the speed of recovery afterward, with Republican-governed counties returning to baseline faster. Bottom: average weekly radius of gyration by rurality level; rural counties maintain consistently higher mobility than urban counties throughout both the restriction and post-restriction periods.}
    \label{weekly_county_rural}
\end{figure}

We see that if, on the one hand, party seems not to be important for the weekly radius of gyration, the rurality level is, with people living in rural counties tending to move more than urban residents throughout the observation window (Fig.~\ref{weekly_county_rural}). Notably, the decline in mobility begins visibly before the mandatory restriction period: California's non-mandatory recommendation to stay home on March 11, 2020 (red dotted line) \cite{hale2021global} already precedes a drop in radius of gyration, which continues and steepens once the mandatory order takes effect a few days later. The decline in radius of gyration was already underway before the statewide mandatory order. This timing is consistent with voluntary behaviour, local measures, work/school measures, or other contemporaneous responses, for both rural and urban populations and across governing parties alike. While our analysis cannot distinguish among these explanations our results give a snapshot of the differences in observed mobility across the restriction windows in California, and highlight the differences in the spatial extent, location discovery, visitation concentration, and trajectory entropy patterns%

    \section*{Discussion}
        \label{discussion}
\begin{figure}[h!]
    \centering
    \includegraphics[width=1\textwidth]{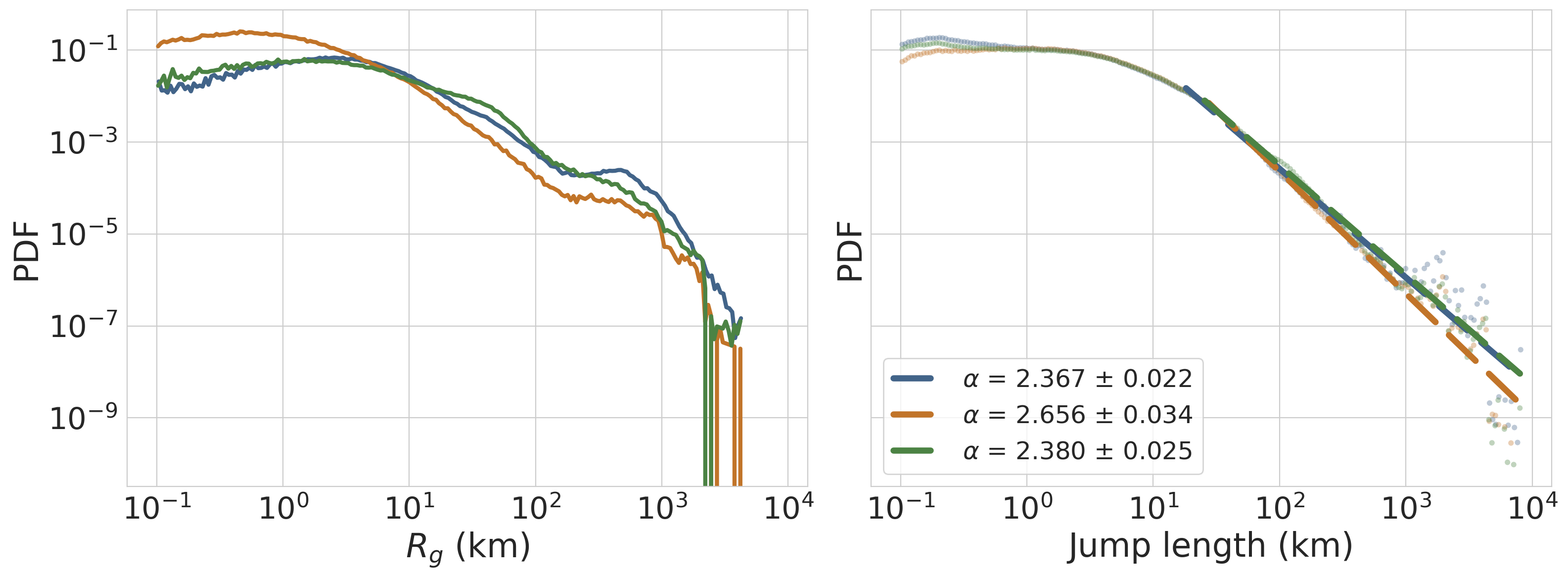}
      \caption{{\bf Changes in mobility across the pandemic phases of social distancing, Massachusetts}. Plot on the left corresponds to the {\it distribution of radius of gyration}, where we can see that the restriction period is characterised by a shift from medium to short-distance mobility, while long-distance trips remain largely unchanged. Plot on the right corresponds to the {\it distribution of jump lengths} between successive stop points. The three distributions overlap closel with differences emerging only in the tail.}
\label{radg_and_dist}
\end{figure}

\begin{figure}[h!]
    \centering
    \includegraphics[width=1\linewidth]{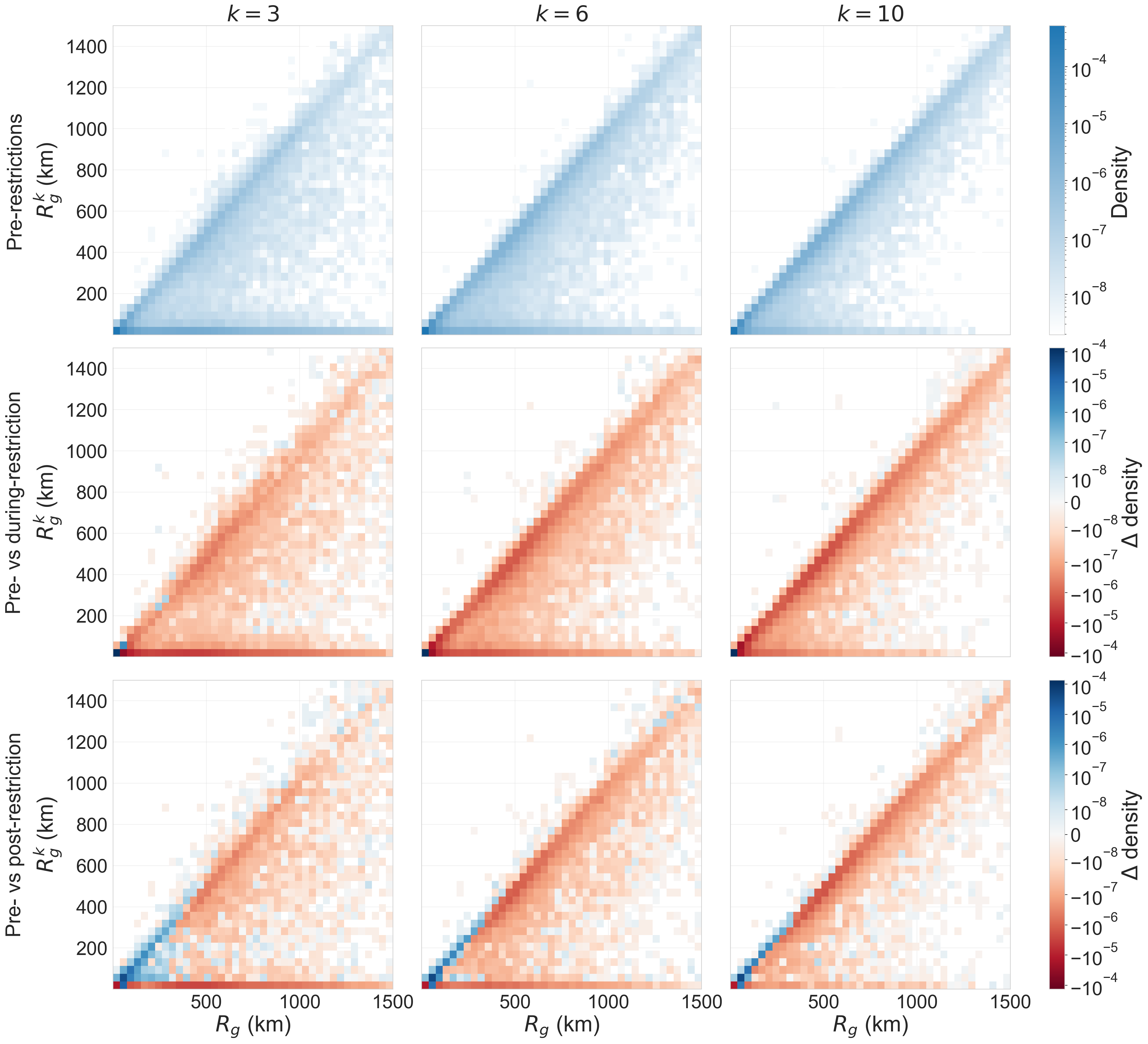}
    \caption{{\bf Change in the joint $R_g$--$R_g^{k}$ distribution relative to the pre-restriction baseline (Massachusetts).} The top row reproduces the pre-restriction  joint density as a reference. The middle and bottom rows show, on the identical $(R_g, R_g^{k})$ bin grid, the absolute density difference between the during-restriction and post-restriction  periods and that same baseline, respectively. Panels show $k = 3, 6, 10$ (left to right). Blue indicates a gain in density relative to the baseline, red a loss. During restriction, density shifts almost uniformly out of the bulk of the $(R_g, R_g^{k})$ plane into the near-zero corner (blue), at the expense of every larger-$R_g$ region (red), consistent with the top-$k$ locations increasingly accounting for the full range of an individual's movement. Post-restriction, the pattern is not a simple reversion to baseline: density along and near the diagonal $R_g^{k} \approx R_g$ increases at small-to-moderate $R_g$ (blue), while the largest-$R_g$ region remains suppressed relative to baseline (red), indicating a rebound concentrated in shorter-range, habitual mobility, with long-distance travel still lagging behind pre-restriction levels -- though this rebound is comparatively more modest than in California (Fig.~\ref{returner_explorers_california}), consistent with the near-zero band staying closer to its baseline share even by September.}    \label{returner_explorers_massachusetts}
\end{figure}

\begin{figure}[h!]
    \centering
    \includegraphics[width=1\textwidth]{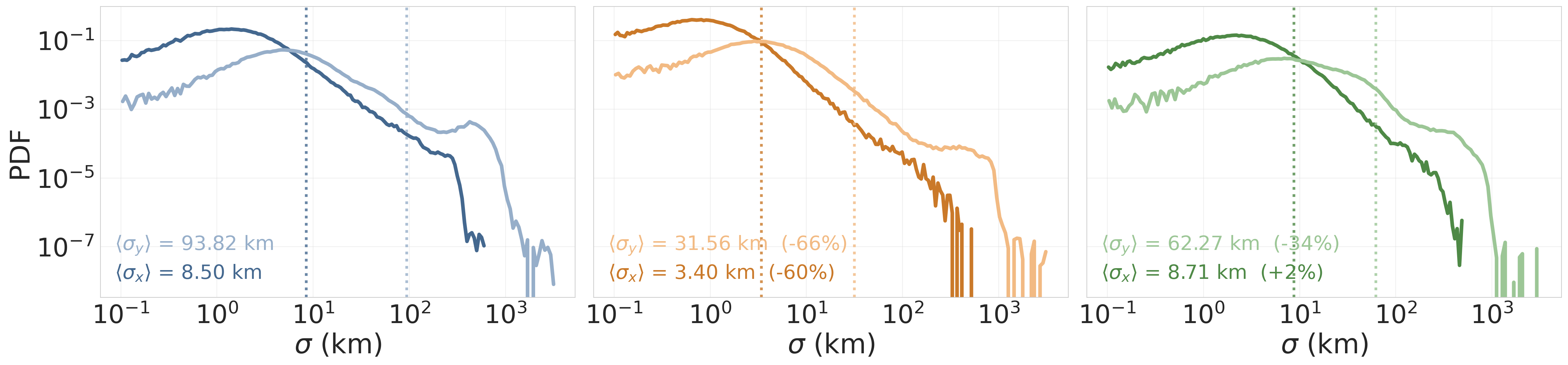}
    
    \caption{{\bf Spread of trajectories along the principal axes, Massachusetts}. Distribution across users of the standard deviations $\sigma_{x'}$  and $\sigma_{y'}$ of stop-point positions along each trajectory's principal axes, in kilometres. Both mean values drop sharply during restriction relative to the pre-restriction period ($\langle\sigma_{x'}\rangle$: $8.54 \to 3.39$~km, $-60\%$; $\langle\sigma_{y'}\rangle$: $93.36 \to 31.54$~km, $-66\%$), consistent with the spatial contraction shown in Fig.~\ref{radg_and_dist}. In the post-restriction period, $\langle\sigma_{x'}\rangle$ recovers close to its pre-restriction value ($8.73$~km, $+2\%$), while $\langle\sigma_{y'}\rangle$ remains reduced ($62.24$~km, $-33\%$), indicating incomplete and asymmetric recovery across the two principal axes.}
    \label{sigma_massachusetts}
\end{figure}

In this work, we investigated whether the changes in mobility observed during the COVID-19 restrictions corresponded to a simple reduction in the spatial scale of movement or to a more fundamental reorganization of individual mobility patterns. 
Understanding this distinction is of potential interest to advance epidemiological models that relies on mobility-derived proxies for potential transmission events. 
In particular, some mobility-based epidemic models ~\cite{Gatto2020, Bertuzzo2020, Aleta2020, Arenas2020} represent behavioural changes primarily by rescaling baseline movement volumes or distances. If these features together with exploration and visitation patterns also change, such representations may omit features of mobility that could be relevant to the model's calibration. 
The change in the structure of these mobility patterns might be reflected on more complex modification of pre-restriction transmission patterns. 
By combining geometric, visitation, and information-theoretic measures across three pandemic phases in two US regions, we characterize several complementary dimensions of individual mobility, and how these changed across the population.
Overall, our results suggest that the restriction-phase window was characterized not only by a reduced spatial extent, but also by a limited location discovery and a greater concentration of visits among frequently visited locations. Moreover, the post-restriction period we consider did not intermediately fully recover the mobility patterns observed in the pre-restriction period, as previously shown \cite{SantanaChanges, Smolyak2021}.
Considering the radius of gyration and jump distances, we observe that the change across these metrics was not a uniform rescaling of all movement distances (Fig.~\ref{radg_and_dist_california} for CA, Fig.~\ref{radg_and_dist} for MA). Short- and medium-range jump-length distributions remained broadly similar across phases, whereas the fitted long-distance tail was modestly lighter during the restriction-phase window and heavier in the later period.
For both California and Massachussetts the radius of gyration concentrated during the restriction on ranges below 5 km. These patterns are stable under $t_{burst}$ change to 8,24 h (Tabs. [\ref{tab:rg_distance_crossing},\ref{tab:distance_powerlaw}]).
In California, post-period density in Fig.~\ref{returner_explorers_california} exceeds the pre-period baseline along and near the diagonal $R_g^{(k)} \approx R_g$ at small-to-moderate $R_g$, while remaining below baseline at the largest $R_g$ -- an asymmetric, not uniform, excess. Recovery in Massachusetts (Fig.~\ref{returner_explorers_massachusetts}, Tab.~\ref{tab:rg_krg_percentile}) was comparatively more incomplete, with the near-zero-$R_g^{(k)}$ concentration remaining closer to its restriction-phase level than California's. We therefore describe the later-period patterns as heterogeneous rather than as a uniform return to baseline. This pattern is stable under changes of $t_{burst}$. The same pattern is observed in the spread of trajectories along their principal axes. In California, both axes recover to values comparable to those observed before the restrictions, whereas in Massachusetts the recovery differs between the two axes (Figs.~\ref{sigma}, \ref{sigma_massachusetts}, and Table~\ref{tab:sigmaxy_pct_change}). This asymmetric difference between post- and pre-period values is also observed at $t^{burst}=8,24$~h (Table~\ref{tab:sigmaxy_pct_change}), suggesting that it does not depend on the choice of temporal resolution. After trajectories were standardized for scale and orientation, their aggregate geometric form remained broadly similar across phases. Observed mobility became more concentrated around frequently visited locations (Figs. \ref{Gonzalez_california},\ref{visitation_increase-frequency_california}). 

\begin{figure}[h!]
    \includegraphics[width=1\textwidth]{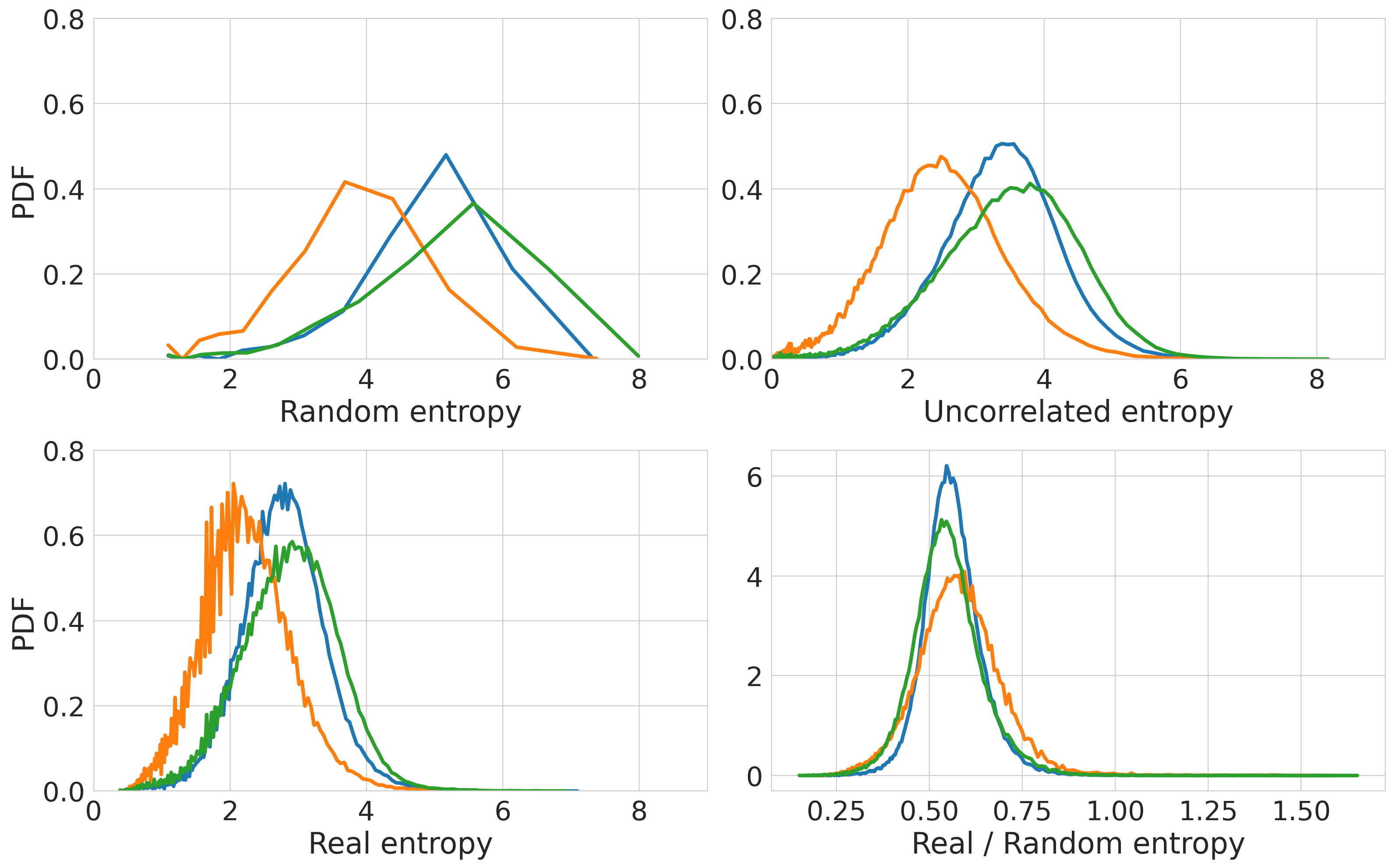}
    \caption{{\bf Entropy measures of individual trajectories, Massachusetts}. Distributions across users of random entropy $H_{\text{random}}$ (top left), uncorrelated entropy $H_{\text{uncorrelated}}$ (top right), and real entropy $H_{\text{real}}$ (bottom left); as in California (Fig.~\ref{entropic_measures_california}). All three entropy measures shift toward lower values during the restriction period, indicating predictability of individual mobility trajectories. Bottom right: distribution of the per-user ratio $H_{\text{real}}/H_{\text{random}}$, which shifts toward higher values during restriction relative to the pre- and post-restriction periods. Corresponding Kolmogorov-Smirnov statistics are reported in Table~\ref{tab:ks_entropy}.}
    \label{entropic_measures_massachusetts}
\end{figure}

The reduction in trajectory entropy during the restriction period (Figs.~\ref{entropic_measures_california}, \ref{entropic_measures_massachusetts}; Table~\ref{tab:entropy_peak}) appears to be driven primarily by the smaller number of distinct locations visited, rather than by changes in the temporal organization of mobility. Although all entropy measures decrease during restrictions, the ratio between real and random entropy seems to move closer to one, indicating that the reduction in entropy is largely explained by the decrease in the number of visited locations. This observation is further supported by the Kolmogorov-Smirnov statistics reported in Table~\ref{tab:ks_entropy} where distributions are compared quantitatively. Previous studies have shown that the high predictability of human mobility mainly arises from the limited number of frequently visited locations rather than from highly repetitive movement patterns~\cite{entropic_measures,Lu2013}. Our results suggest that pandemic restrictions reinforced this same mechanism by reducing the number of locations people visited, rather than by making their mobility routines substantially more repetitive.

\begin{figure}[h!]
    \centering
    \includegraphics[width=1\textwidth]{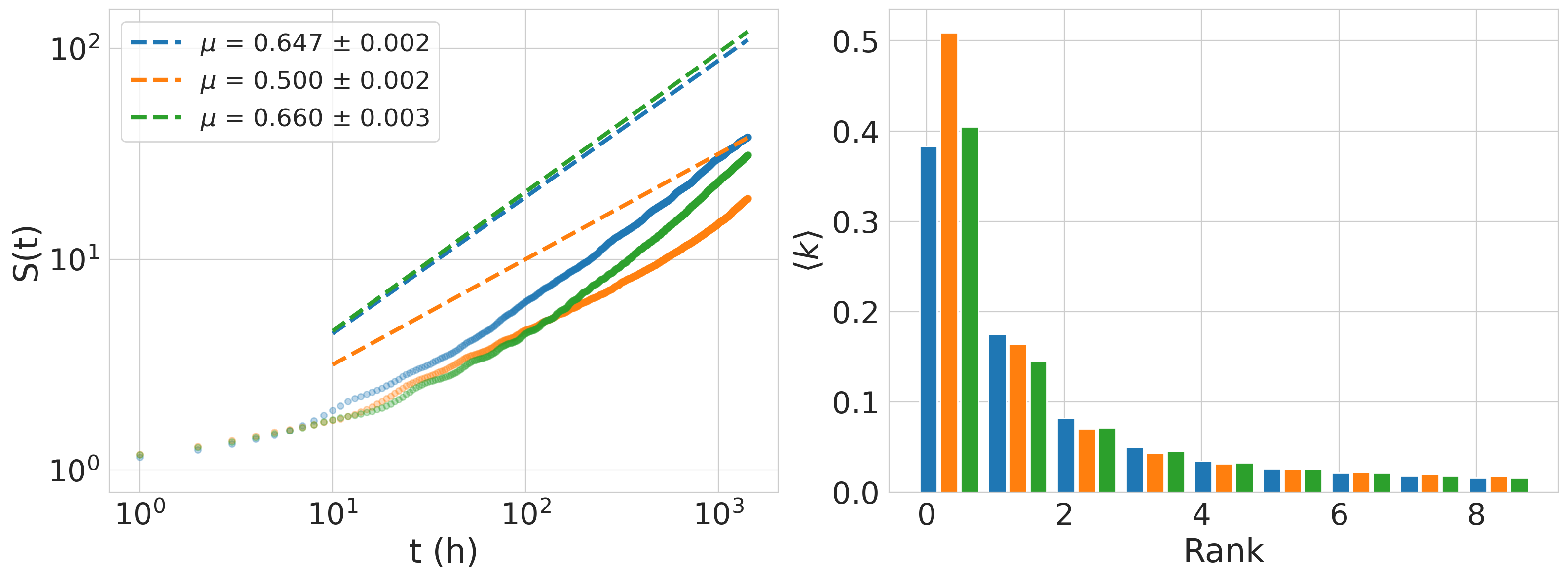}
    \caption{{\bf Exploration rate and visitation-frequency ranking, Massachusetts}. Left: average number of distinct stop points discovered over time, $S(t)$, with power-law fits $S(t) \sim t^{\mu}$ yielding $\mu_{\text{pre}} = 0.647 \pm 0.002$, $\mu_{\text{during}} = 0.500 \pm 0.002$, and $\mu_{\text{post}} = 0.660 \pm 0.003$; as in California (Fig.~\ref{visitation_increase-frequency_california}), the exploration process during restriction approaches diffusive behaviour ($\mu \approx 0.5$). Right: normalised visit frequency by location rank for the ten most-visited places per user; the two most-visited locations account for 67\% of visits during restriction, against approximately 55\% in the pre- and post-restriction periods.}
    \label{visitation_increase-frequency_massachusetts}
\end{figure}

The discovery-rate and visitation-concentration analyses support this picture (Fig.~\ref{visitation_increase-frequency_california}, Fig.~\ref{visitation_increase-frequency_massachusetts}; Tables~\ref{tab:st_exponent}, \ref{tab:frequency_top2}) and by the joint distribution of the radius of gyration and the $k$-radius of gyration (Fig.~\ref{returner_explorers_california}, Fig.~\ref{returner_explorers_massachusetts}; Table~\ref{tab:rg_krg_percentile}): people explored fewer places and concentrated more visits on their existing habitual locations, becoming more returners than explorers \cite{pappalardo_returners}.

The discovery-rate exponent drops from super-diffusive to near-diffusive during restrictions and recovers afterward, matching scaling laws for exploration and visitation frequency found outside the pandemic \cite{scaling_visitation}. The fitted discovery curves remained compatible with the same broad scaling framework despite the large behavioural disruption. During the restriction-phase window, geographic range decreased and mobility became more concentrated among frequently visited locations, rather than making individual routines more rigid.
Notably, we believe this is a distinction easy to elide when only aggregate mobility volume is reported, as in several early-pandemic mobility indices \cite{GoogleMobility, AppleMobility}.

In the exploratory California stratification, users assigned to rural counties had higher absolute radii of gyration than users assigned to urban counties throughout the observation period. Differences between county political groups were smaller and were not statistically distinguishable during the restriction-phase window (Fig.~\ref{weekly_county_rural}). Differences from previous county-level studies may partly reflect the use of different mobility outcomes, such as maximum distance travelled, points-of-interest visits, or stay-at-home measures, which are not directly equivalent to the duration-weighted radius of gyration used here \cite{Kim2021,Allcott2020,Grossman2020}. We stress that radius of gyration captures a different dimension of mobility from stay-at-home rates or points-of-interest counts, so direct comparisons across studies should be made cautiously. Moreover, this analysis was conducted only for California, and uses unadjusted county-level classifications in an opt-in mobile-app sample. The results should therefore be interpreted as descriptive and ecological, not as individual-level or causal effects of rurality or political context \cite{Wesolowski2013, Zhao2016, Li2024}.

Our research has several limitations that should guide how these results are interpreted, and serve as follow-up work that could improve on them. On the data side, the Cuebiq data is subject to demographic sampling biases common to opt-in mobile location data \cite{Wesolowski2013, Zhao2016, Li2024}, which we have not attempted to re-weight or correct for; if smartphone ownership or app usage co-varies with rurality or political geography, this could confound the heterogeneity results in Section~\ref{weekly_county_rural}. 
Moreover, the number of retained observations per user differed substantially across phases. Because estimates of distinct locations, discovery rates, and Lempel–Ziv entropy depend on trajectory length and sampling intensity, part of the observed differences may reflect finite-sequence or observation-density effects. 
On the methodological side, the burst-filter threshold $t^{burst}$ meaningfully affects the absolute value of the jump-length tail exponent, even though the qualitative ordering across pandemic phases is preserved at every threshold tested in both regions (Table~\ref{tab:distance_powerlaw}); a more principled, data-driven choice of this threshold, rather than a fixed set of candidate values, would strengthen future comparisons across studies. More broadly, the first pandemic wave constituted an unusually large and geographically widespread disruption to routine mobility, sudden, large-scale, and geographically synchronised in ways that differ from ordinary mobility perturbations, and it remains an open question how much of what we report here (the rebound signature in particular) is specific to this event versus a general feature of large exogenous shocks to routine mobility. Finally, our heterogeneity analysis is descriptive and county-level; individual-level co-variates (income, occupation, essential-worker status) were not available in this dataset, and their inclusion would help disentangle whether the rurality and party effects we report are themselves proxies for other socioeconomic gradients \cite{Weill2020}.

In this work we treat the pandemic period uniformly through three phases, a deliberate simplification. In fact, no single date can fully capture the onset of behavioural change: public information, local restrictions, and statewide policies occurred at different times. The three-phase segmentation should therefore be interpreted as a descriptive simplification rather than a causal policy design. Different degrees and timing of restriction severity were, however, present in both California and Massachusetts: \cite{hale2021global} shows that individual counties in both states adopted restrictions with different timings and different levels of severity, a county-level distinction our state-level, three-phase segmentation avoids. Notably, the median radius of gyration in Fig.~\ref{weekly_county_rural} already begins declining approximately 2 weeks before 15~March across all four subgroups (rural/urban, Democratic/Republican), i.e. before the mandatory restrictions we use to delimit the restriction phase were in place. The decline began before the statewide mandatory restriction date, showing that the phase comparison cannot isolate the effect of formal policy. Possible contributors include voluntary responses to perceived risk, local policies, workplace and school decisions, and other contemporaneous changes.
The epidemiological relevance of these findings is indirect. The GPS trajectories characterize movement locations, spatial extent, and dwell patterns, but do not measure interpersonal co-presence, contact intensity, infection, or transmission. We therefore do not infer that the observed mobility changes produced corresponding changes in transmission risk. Rather, the results identify mobility features that may need to be re-estimated when mobility data are used as inputs to epidemic models. Whether incorporating these features improves estimates of contact patterns, transmission, or epidemic forecasts would require direct model evaluation using epidemiological outcomes.

    \section*{Acknowledgments}
    B.L. and S.C. acknowledge partial financial support from the following projects: Horizon Europe Programme, grants \#10112- 0237-ELIAS and \#101120763-TANGO. Funded by the European Union. Views and opinions expressed are however those of the author(s) only and do not necessarily reflect those of the European Union or the European Health and Digital Executive Agency (HaDEA). Neither the European Union nor the granting authority can be held responsible for them. V.dA. and M.D.D. acknowledge partial financial support from the INFN grant “LINCOLN”.
    
    \section*{Data Availability}
    The code to reproduce the plots with preprocessed data and the algorithms used to process the raw mobility data are available in \cite{amaduzzi2020humanmobilitycovid}

    \section*{Declarations}
    The authors declare no conflict of competing interests.

    \newpage
    \section*{Supplementary Information}
       \label{supplementary_information}

\subsection*{The Massachusetts case study}

In this section, we reproduce the analyses presented in the main text using the Massachusetts dataset. As for the case of California, trajectories are constructed using a time burst filter threshold of $t_{burst}=1$. A summary of the statistics for the dataset across the different periods of study are reported in Table~\ref{tabPeopleCal}.

{
\renewcommand{\arraystretch}{1.5} 
\setlength{\tabcolsep}{12pt}     
\begin{table}[H]
    \centering
    \begin{tabular}{l l r r r}
        \rowcolor{gray!5}
        Period & Dates & $N_m$ & $T_m$ & $\langle$ $T_m$ $\rangle_p$\\
        \rowcolor{figblue!20}
        Pre- restrictions & 15 January - 15 March    & 181778   & 28782953  & 158\\
        \rowcolor{figorange!20}
        During- restrictions & 15 March - 15 May    & 140913   & 12701431  & 90\\
        \rowcolor{figgreen!20}
        Post- restrictions & 15 May - 30 September    & 141285   & 32096311 & 227 \\
    \end{tabular}
    \caption{Main information about the Massachusetts preprocessed dataset with columns $N_c$ number of users, number of total points $T_m$ and average points per each trajectory $\langle T_m \rangle_p$}
    \label{tabPeopleMA}
\end{table}
}

\begin{figure}[h!]
    \centering
    \includegraphics[width=1\textwidth]{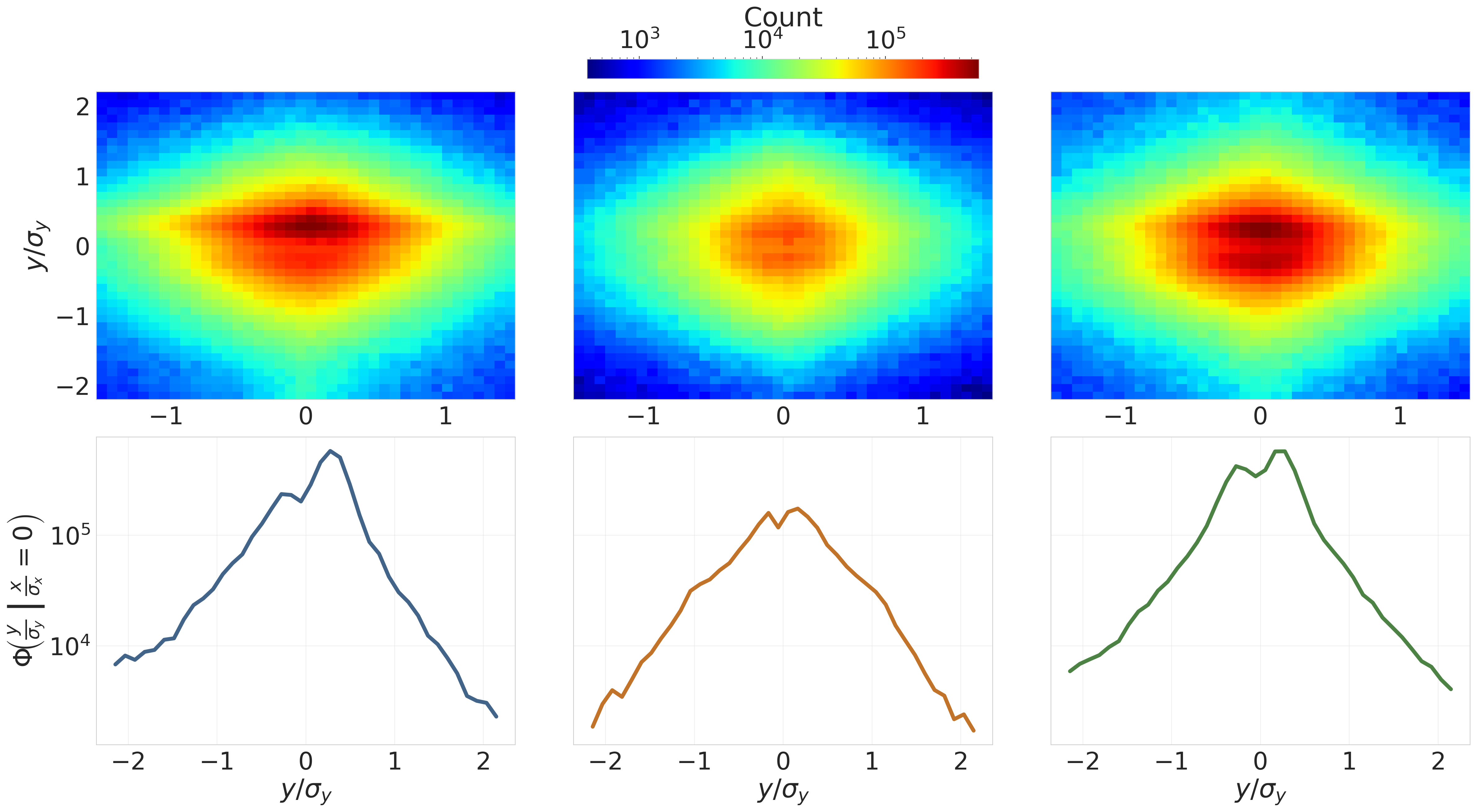}
    \caption{{\bf Trajectory shape in the intrinsic reference frame, Massachusetts}. Top: two-dimensional distribution of stop-point positions in each trajectory's principal-axis frame, rescaled by the standard deviations of the rotated coordinates. Bottom: conditional probability density along $\hat{y}$ at $\hat{x}=0$. As for California (Fig.~\ref{Gonzalez_california}), the overall trajectory geometry remains stable across periods. During the restriction phase, however, the density becomes more concentrated around the origin, emerging a reduced spatial extent of individual mobility. This contraction is quantified in Fig.~\ref{sigma_massachusetts}.}
    \label{Gonzalez_massachusetts}
\end{figure}

Overall, the Massachusetts analysis reveals the same qualitative mobility patterns observed in California. The restriction period is characterized by a contraction of individual mobility, reduced exploration, increased concentration around habitual locations, and lower trajectory entropy. The consistency of these findings across two geographically distinct regions might suggests that the observed behavioural changes reflect general responses to the pandemic restrictions.

\subsection*{Sensitivity analysis}

To evaluate the robustness of our findings, we recompute every quantity reported in the main text (see the Results Section) at $t^{burst} = 1, 8,$ and $24$ hours. The tables below report the underlying numbers directly, so that each qualitative claim in the main text can be checked against its value at every threshold tested.

{
\renewcommand{\arraystretch}{1.3}
\setlength{\tabcolsep}{8pt}
\begin{table}[t!]
    \centering
    \begin{tabular}{l l l r r }
        \hline
        Region & $t^{burst}$ & Comparison & $r_g^{*}$ (km) & Excess mass ($r_g<r_g^{*}$) \\
        \hline
\multirow{2}{*}{California} & \multirow{2}{*}{1 h}
    & during -- pre   & 5.40 & 0.2662 \\
    &   & during -- post  & 11.63 & 0.2901 \\
\hline
\multirow{2}{*}{Massachusetts} & \multirow{2}{*}{1 h}
    & during -- pre   & 4.79 & 0.3318  \\
    &   & during -- post  & 7.72 & 0.3752 \\
\hline
\multirow{2}{*}{California} & \multirow{2}{*}{8 h}
    & during -- pre   & 5.55 & 0.2494  \\
    &   & during -- post  & 11.16 & 0.2810 \\
\hline
\multirow{2}{*}{Massachusetts} & \multirow{2}{*}{8 h}
    & during -- pre   & 5.05 & 0.3154 \\
    &   & during -- post  & 8.00 & 0.3660 \\
\hline
\multirow{2}{*}{California} & \multirow{2}{*}{24 h}
    & during -- pre   & 5.17 & 0.2515 \\
    &   & during -- post  & 10.54 & 0.2792 \\
\hline
\multirow{2}{*}{Massachusetts} & \multirow{2}{*}{24 h}
    & during -- pre   & 4.76 & 0.3135 \\
    &   & during -- post  & 7.75 & 0.3653 \\
\hline
    \end{tabular}
    \caption{{\bf Crossing point and redistribution of probability mass in the radius-of-gyration distribution during restriction.} $r_g^{*}$ is the radius at which the during-restriction $R_g$ PDF, elevated at short distances, drops back below the comparison period's PDF. ``Excess mass'' is the signed area between the two PDFs below $r_g^{*}$ respectively (probability mass, since each PDF integrates to 1 over km) -- i.e. the density gained at short trips and lost at long trips during restriction relative to the comparison period. The near-equal magnitude of excess and deficit mass at every $t^{burst}$ indicates medium/long-distance mobility is redistributed onto short trips rather than lost outright.
    This confirms the $\sim5$~km contraction scale reported in the main text (Fig.~\ref{radg_and_dist_california}) and shows it is stable across $t^{burst} \in \{1,8,24\}$~h and across both regions.
    }
    \label{tab:rg_distance_crossing}
\end{table}
}

{
\renewcommand{\arraystretch}{1.3}
\setlength{\tabcolsep}{8pt}
\begin{table}[t!]
    \centering
    \begin{tabular}{l l l r r r r}
        \hline
        Region & $t^{burst}$ & Period & $\alpha$ & $\sigma_\alpha$ & $x_{min}$ (km) & Tail fraction \\
        \hline
\multirow{3}{*}{California} & \multirow{3}{*}{1 h}
    & pre     & 2.331 & 0.025 & 29.97 & 13.8\% \\
    &   & during  & 2.370 & 0.024 & 29.84 & 12.5\% \\
    &   & post    & 2.233 & 0.023 & 26.57 & 15.6\% \\
\hline
\multirow{3}{*}{Massachusetts} & \multirow{3}{*}{1 h}
    & pre     & 2.367 & 0.022 & 24.51 & 13.4\% \\
    &   & during  & 2.656 & 0.034 & 29.86 & 10.3\% \\
    &   & post    & 2.380 & 0.025 & 20.32 & 19.3\% \\
\hline
\multirow{3}{*}{California} & \multirow{3}{*}{8 h}
    & pre     & 2.307 & 0.023 & 29.47 & 15.9\% \\
    &   & during  & 2.377 & 0.024 & 29.57 & 15.5\% \\
    &   & post    & 2.196 & 0.020 & 29.77 & 18.0\% \\
\hline
\multirow{3}{*}{Massachusetts} & \multirow{3}{*}{8 h}
    & pre     & 2.483 & 0.026 & 27.12 & 16.2\% \\
    &   & during  & 2.685 & 0.033 & 29.51 & 12.8\% \\
    &   & post    & 2.207 & 0.016 & 19.20 & 26.9\% \\
\hline
\multirow{3}{*}{California} & \multirow{3}{*}{24 h}
    & pre     & 2.114 & 0.019 & 29.77 & 17.6\% \\
    &   & during  & 2.229 & 0.022 & 29.90 & 15.0\% \\
    &   & post    & 1.986 & 0.014 & 23.16 & 24.3\% \\
\hline
\multirow{3}{*}{Massachusetts} & \multirow{3}{*}{24 h}
    & pre     & 2.204 & 0.018 & 22.89 & 21.4\% \\
    &   & during  & 2.396 & 0.025 & 23.91 & 16.4\% \\
    &   & post    & 2.007 & 0.012 & 16.37 & 32.6\% \\
\hline
    \end{tabular}
    \caption{{\bf Power-law fit of the inter-stop jump-length distribution across pandemic phases.} $\alpha$ is the Clauset--Shalizi--Newman maximum-likelihood power-law exponent, $\sigma_\alpha$ its standard error, $x_{min}$ the fitted lower cutoff, and ``Tail fraction'' the share of jumps at or above $x_{min}$, for each period and burst-filter threshold $t^{burst}$. The exponent steepens (larger $\alpha$) and the tail fraction shrinks during restriction relative to pre/post, consistently across every $t^{burst}$ tested.
    These fits reproduce the exponents quoted in the main text for $t^{burst}=1$~h (Fig.~\ref{radg_and_dist_california}) and $t^{burst}=8$~h (Fig.~\ref{radg and dist 8}), and extend the comparison to $t^{burst}=24$~h; the during-restriction exponent is the largest and the post-restriction exponent the smallest at every threshold, confirming that the ordering — not the absolute value — is the robust finding (see Discussion Section).
    }
    \label{tab:distance_powerlaw}
\end{table}
}

{
\renewcommand{\arraystretch}{1.3}
\setlength{\tabcolsep}{8pt}
\begin{table}[t!]
    \centering
    \begin{tabular}{l l l r r r}
        \hline
        Region & $t^{burst}$ & Period & $\mu$ & $\sigma_\mu$ & $r$ \\
        \hline
\multirow{3}{*}{California} & \multirow{3}{*}{1 h}
    & pre     & 0.672 & 0.0017 & 0.9955 \\
    &   & during  & 0.529 & 0.0018 & 0.9918 \\
    &   & post    & 0.672 & 0.0027 & 0.9885 \\
\hline
\multirow{3}{*}{Massachusetts} & \multirow{3}{*}{1 h}
    & pre     & 0.647 & 0.0018 & 0.9947 \\
    &   & during  & 0.500 & 0.0018 & 0.9909 \\
    &   & post    & 0.660 & 0.0030 & 0.9860 \\
\hline
\multirow{3}{*}{California} & \multirow{3}{*}{8 h}
    & pre     & 0.649 & 0.0033 & 0.9977 \\
    &   & during  & 0.548 & 0.0041 & 0.9951 \\
    &   & post    & 0.661 & 0.0057 & 0.9936 \\
\hline
\multirow{3}{*}{Massachusetts} & \multirow{3}{*}{8 h}
    & pre     & 0.628 & 0.0039 & 0.9966 \\
    &   & during  & 0.518 & 0.0040 & 0.9948 \\
    &   & post    & 0.653 & 0.0063 & 0.9919 \\
\hline
\multirow{3}{*}{California} & \multirow{3}{*}{24 h}
    & pre     & 0.678 & 0.0033 & 0.9993 \\
    &   & during  & 0.586 & 0.0061 & 0.9969 \\
    &   & post    & 0.678 & 0.0070 & 0.9970 \\
\hline
\multirow{3}{*}{Massachusetts} & \multirow{3}{*}{24 h}
    & pre     & 0.659 & 0.0038 & 0.9991 \\
    &   & during  & 0.558 & 0.0060 & 0.9967 \\
    &   & post    & 0.667 & 0.0083 & 0.9957 \\
\hline
    \end{tabular}
    \caption{{\bf Exploration exponent of the $S(t)$ curve across pandemic phases.} $\mu$ is the power-law exponent of the average number of distinct locations visited by time $t$, $S(t) \sim t^{\mu}$, fitted by OLS in log-log space; $\sigma_\mu$ its standard error and $r$ the fit's correlation coefficient. $\mu$ drops markedly during restriction at every burst-filter threshold $t^{burst}$, indicating a slower rate of discovery of new locations, and recovers afterwards.
    These values confirm the $\mu$ estimates reported in the main text (Fig.~\ref{visitation_increase-frequency_california}) and their Massachusetts counterpart (Fig.~\ref{visitation_increase-frequency_massachusetts}) at $t^{burst}=1$~h, and show the drop to near-diffusive behaviour during restriction, and recovery afterward, at $t^{burst}=8$ and $24$~h as well.
    }
    \label{tab:st_exponent}
\end{table}
}

{
\renewcommand{\arraystretch}{1.3}
\setlength{\tabcolsep}{8pt}
\begin{table}[t!]
    \centering
    \begin{tabular}{l l l r r r}
        \hline
        Region & $t^{burst}$ & Period & Rank 1 (\%) & Rank 2 (\%) & Rank 1+2 (\%) \\
        \hline
\multirow{3}{*}{California} & \multirow{3}{*}{1 h}
    & pre     & 36.10 & 16.82 & 52.92 \\
    &   & during  & 46.96 & 16.00 & 62.96 \\
    &   & post    & 39.46 & 14.37 & 53.83 \\
\hline
\multirow{3}{*}{Massachusetts} & \multirow{3}{*}{1 h}
    & pre     & 38.26 & 17.47 & 55.74 \\
    &   & during  & 50.90 & 16.39 & 67.29 \\
    &   & post    & 40.49 & 14.51 & 55.00 \\
\hline
\multirow{3}{*}{California} & \multirow{3}{*}{8 h}
    & pre     & 35.66 & 18.31 & 53.98 \\
    &   & during  & 44.93 & 17.22 & 62.14 \\
    &   & post    & 36.99 & 15.55 & 52.54 \\
\hline
\multirow{3}{*}{Massachusetts} & \multirow{3}{*}{8 h}
    & pre     & 37.08 & 19.16 & 56.24 \\
    &   & during  & 48.73 & 17.58 & 66.31 \\
    &   & post    & 37.83 & 15.63 & 53.45 \\
\hline
\multirow{3}{*}{California} & \multirow{3}{*}{24 h}
    & pre     & 35.50 & 17.17 & 52.68 \\
    &   & during  & 45.63 & 16.55 & 62.18 \\
    &   & post    & 37.63 & 14.58 & 52.21 \\
\hline
\multirow{3}{*}{Massachusetts} & \multirow{3}{*}{24 h}
    & pre     & 37.35 & 17.98 & 55.33 \\
    &   & during  & 49.68 & 16.85 & 66.53 \\
    &   & post    & 38.86 & 14.49 & 53.34 \\
\hline
    \end{tabular}
    \caption{{\bf Share of total visitation captured by the two most-visited locations, across pandemic phases.} Rank~1/Rank~2 are the mean fraction of a user's visits spent at their most- and second-most-visited location respectively (Eq.~\ref{k_avg}); Rank~1+2 is their sum. The combined share rises sharply during restriction at every burst-filter threshold $t^{burst}$, then relaxes back towards the pre-restriction level.
    These confirm the concentration figures quoted in the main text: 62.96\%/52.92\%/53.83\% for California and 67.29\%/55.74\%/55.00\% for Massachusetts at $t^{burst}=1$~h (during/pre/post); the same pattern holds at $8$ and $24$~h.
    }
    \label{tab:frequency_top2}
\end{table}
}

{
\renewcommand{\arraystretch}{1.3}
\setlength{\tabcolsep}{8pt}
\begin{table}[t!]
    \centering
    \begin{tabular}{l l l r r r}
        \hline
        Region & $t^{burst}$ & Period & $P_{95}(R_g \mid k=3) (km)$ & $P_{95}(R_g \mid k=6) (km)$ & $P_{95}(R_g \mid k=10) (km)$ \\
        \hline
\multirow{3}{*}{California} & \multirow{3}{*}{1 h}
    & pre     & 272.3 & 96.0 & 44.7 \\
    &   & during  & 41.4 & 29.6 & 26.7 \\
    &   & post    & 221.2 & 97.2 & 54.1 \\
\hline
\multirow{3}{*}{Massachusetts} & \multirow{3}{*}{1 h}
    & pre     & 421.3 & 78.5 & 37.7 \\
    &   & during  & 28.2 & 23.2 & 21.7 \\
    &   & post    & 128.2 & 52.4 & 38.3 \\
\hline
\multirow{3}{*}{California} & \multirow{3}{*}{8 h}
    & pre     & 289.7 & 106.2 & 49.9 \\
    &   & during  & 42.4 & 29.9 & 26.7 \\
    &   & post    & 240.7 & 116.3 & 64.7 \\
\hline
\multirow{3}{*}{Massachusetts} & \multirow{3}{*}{8 h}
    & pre     & 420.4 & 112.5 & 39.1 \\
    &   & during  & 28.9 & 23.3 & 21.5 \\
    &   & post    & 149.6 & 60.4 & 40.6 \\
\hline
\multirow{3}{*}{California} & \multirow{3}{*}{24 h}
    & pre     & 272.4 & 109.3 & 48.6 \\
    &   & during  & 41.9 & 29.8 & 26.3 \\
    &   & post    & 237.2 & 122.6 & 68.9 \\
\hline
\multirow{3}{*}{Massachusetts} & \multirow{3}{*}{24 h}
    & pre     & 392.2 & 132.9 & 36.8 \\
    &   & during  & 29.0 & 22.7 & 20.6 \\
    &   & post    & 142.6 & 66.9 & 42.6 \\
\hline
    \end{tabular}
    \caption{{\bf Unexplained mobility range for users whose $k$-radius of gyration is near zero.} $P_{95}(R_g \mid k)$ (km) is the 95th percentile of the full radius of gyration $R_g$ among users whose $k$-radius of gyration falls in the first (near-zero) histogram bin -- i.e. users whose top-$k$ habitual locations alone would suggest almost no mobility. A large $P_{95}$ means a non-trivial fraction of overall mobility is not explained by the $k$ most-visited locations. The lower $P_{95}$ during restriction, consistent across $t^{burst}$, shows habitual locations account for a larger share of total mobility under restriction.
    This quantifies, across thresholds, the qualitative pattern shown in Fig.~\ref{returner_explorers_california}: the near-zero band in the joint ($R_g$, $R_g^{k}$) distribution, users whose top-$k$ locations do not explain their overall spread, shrinks sharply during restriction and re-expands afterward.
    }
    \label{tab:rg_krg_percentile}
\end{table}
}

{
\renewcommand{\arraystretch}{1.3}
\setlength{\tabcolsep}{8pt}
\begin{table}[t!]
    \centering
    \begin{tabular}{l l l r r r r}
        \hline
        Region & $t^{burst}$ & Period & $\sigma_x$ (km) & $\Delta\sigma_x$ (\%) & $\sigma_y$ (km) & $\Delta\sigma_y$ (\%) \\
        \hline
\multirow{3}{*}{California} & \multirow{3}{*}{1 h}
    & pre     & 13.07 & -- & 77.47 & -- \\
    &   & during  & 7.33 & -44.0\% & 36.50 & -52.9\% \\
    &   & post    & 19.11 & 46.2\% & 92.68 & 19.6\% \\
\hline
\multirow{3}{*}{Massachusetts} & \multirow{3}{*}{1 h}
    & pre     & 8.50 & -- & 93.82 & -- \\
    &   & during  & 3.40 & -60.0\% & 31.56 & -66.4\% \\
    &   & post    & 8.71 & 2.4\% & 62.27 & -33.6\% \\
\hline
\multirow{3}{*}{California} & \multirow{3}{*}{8 h}
    & pre     & 11.90 & -- & 72.62 & -- \\
    &   & during  & 6.20 & -47.9\% & 32.98 & -54.6\% \\
    &   & post    & 16.77 & 41.0\% & 84.69 & 16.6\% \\
\hline
\multirow{3}{*}{Massachusetts} & \multirow{3}{*}{8 h}
    & pre     & 7.83 & -- & 85.76 & -- \\
    &   & during  & 2.96 & -62.2\% & 28.57 & -66.7\% \\
    &   & post    & 7.91 & 1.0\% & 57.45 & -33.0\% \\
\hline
\multirow{3}{*}{California} & \multirow{3}{*}{24 h}
    & pre     & 11.31 & -- & 70.65 & -- \\
    &   & during  & 5.60 & -50.5\% & 31.24 & -55.8\% \\
    &   & post    & 15.82 & 39.9\% & 81.76 & 15.7\% \\
\hline
    \end{tabular}
    \caption{{\bf Trajectory-shape spread in the Gonzalez intrinsic reference frame across pandemic phases.} $\sigma_x$/$\sigma_y$ are the mean standard deviations of stop points along the principal axes of each user's trajectory in its intrinsic (Gonzalez PCA) reference frame; $\Delta\sigma$ is the percent change relative to the pre-restriction period (``--'' for the pre-restriction baseline itself). Both axes contract sharply during restriction and partially recover afterwards, at every burst-filter threshold $t^{burst}$.
    These percentages confirm the contraction and post-restriction overshoot reported for California in the main text (Fig.~\ref{sigma}) and the asymmetric recovery reported for Massachusetts (Fig.~\ref{sigma_massachusetts}), and show both patterns hold at $t^{burst}=8$ and $24$~h.
    }
    \label{tab:sigmaxy_pct_change}
\end{table}
}

{
\renewcommand{\arraystretch}{1.3}
\setlength{\tabcolsep}{8pt}
\begin{table}[t!]
    \centering
    \begin{tabular}{l l l r r r r}
        \hline
        Region & $t^{burst}$ & Period & $\hat H_{\text{random}}$ & $\hat H_{\text{uncorrelated}}$ & $\hat H_{\text{real}}$ & $\hat H_{\text{real/random}}$ \\
        \hline
\multirow{3}{*}{California} & \multirow{3}{*}{1 h}
    & pre     & 5.407 & 3.601 & 3.065 & 0.557 \\
    &   & during  & 4.512 & 2.726 & 2.556 & 0.578 \\
    &   & post    & 5.813 & 3.647 & 2.989 & 0.546 \\
\hline
\multirow{3}{*}{Massachusetts} & \multirow{3}{*}{1 h}
    & pre     & 5.168 & 3.378 & 2.807 & 0.546 \\
    &   & during  & 3.687 & 2.484 & 2.051 & 0.595 \\
    &   & post    & 5.570 & 3.792 & 2.915 & 0.533 \\
\hline
\multirow{3}{*}{California} & \multirow{3}{*}{8 h}
    & pre     & 4.700 & 3.371 & 2.729 & 0.633 \\
    &   & during  & 3.986 & 2.795 & 2.409 & 0.641 \\
    &   & post    & 5.177 & 3.638 & 2.795 & 0.616 \\
\hline
\multirow{3}{*}{Massachusetts} & \multirow{3}{*}{8 h}
    & pre     & 4.500 & 3.133 & 2.596 & 0.634 \\
    &   & during  & 3.836 & 2.398 & 2.097 & 0.614 \\
    &   & post    & 4.987 & 3.598 & 2.789 & 0.615 \\
\hline
\multirow{3}{*}{California} & \multirow{3}{*}{24 h}
    & pre     & 4.036 & 3.219 & 3.072 & 0.692 \\
    &   & during  & 3.463 & 2.573 & 2.141 & 0.668 \\
    &   & post    & 4.537 & 3.445 & 2.792 & 0.652 \\
\hline
\multirow{3}{*}{Massachusetts} & \multirow{3}{*}{24 h}
    & pre     & 3.843 & 2.962 & 2.974 & 0.727 \\
    &   & during  & 3.364 & 2.337 & 1.787 & 0.667 \\
    &   & post    & 4.365 & 3.376 & 2.793 & 0.658 \\
\hline
    \end{tabular}
    \caption{{\bf Peak (modal) value of each entropy distribution across pandemic phases.} $\hat H$ denotes the location of the PDF peak (mode) for each entropy measure (Eqs.~\ref{rdm_entropy}--\ref{real_entropy}), in bits, for each period (pre-restriction: 15 January -- 15 March; during restriction: 15 March -- 15 May; post-restriction: 15 May -- September) and burst-filter threshold $t^{burst}$. The consistent leftward shift of every entropy measure during restriction, recovering afterwards, shows the drop in trajectory complexity is not an artefact of a particular $t^{burst}$ choice.
    The consistent leftward shift at every threshold confirms that the entropy reduction reported in the main text (Fig.~\ref{entropic_measures_california}) is not an artefact of the $t^{burst}=1$~h choice.
    }
    \label{tab:entropy_peak}
\end{table}
}

{
\renewcommand{\arraystretch}{1.3}
\setlength{\tabcolsep}{4pt}
\begin{table}[t!]
    \centering
    \resizebox{\textwidth}{!}{%
    \begin{tabular}{l l l r r r r r r r r r r r r r r r r}
        \hline
        Region & $t^{burst}$ & Period & $r_g^{*}$(vs pre) & $r_g^{*}$(vs post) & $\alpha$ & $\sigma_\alpha$ & $\mu$ & $\sigma_\mu$ & Rank 1+2 (\%) & $P_{95}(k{=}3)$ & $P_{95}(k{=}6)$ & $P_{95}(k{=}10)$ & $\Delta\sigma_x$ (\%) & $\Delta\sigma_y$ (\%) & $\hat H_{\text{rand}}$ & $\hat H_{\text{uncorr}}$ & $\hat H_{\text{real}}$ & $\hat H_{\text{real/rand}}$ \\
        \hline
\multirow{3}{*}{Massachusetts} & \multirow{3}{*}{1 h}
    & pre     & -- & -- & 2.634 & 0.033 & 0.649 & 0.0018 & 55.70 & 424.0 & 80.3 & 37.8 & -- & -- & 5.352 & 3.475 & 2.898 & 0.539 \\
    &   & during  & 4.48 & 7.03 & 2.775 & 0.038 & 0.569 & 0.0027 & 68.34 & 23.5 & 20.8 & 20.2 & -65.0\% & -76.2\% & 3.806 & 2.480 & 1.965 & 0.594 \\
    &   & post    & -- & -- & 2.347 & 0.022 & 0.660 & 0.0030 & 55.00 & 127.7 & 52.0 & 38.2 & 0.3\% & -34.8\% & 5.835 & 3.850 & 2.881 & 0.539 \\
\hline
    \end{tabular}%
    }
    \caption{{\bf Sensitivity of every mobility metric to the pre/during-restriction boundary date.} Same metrics and periods as the main sensitivity tables (Tables~\ref{tab:rg_distance_crossing}--\ref{tab:entropy_peak}), recomputed with the pre/during-restriction boundary moved from the standard 15 March to 23 March 2020 -- the date the recommendation not to leave home was broadcast -- instead of the standard boundary, for Massachusetts at $t^{burst}=1$~h. $r_g^{*}$ is the radius-of-gyration PDF crossing point (during vs. the labeled comparison period); $\alpha$/$\sigma_\alpha$ the jump-length power-law exponent and its standard error; $\mu$/$\sigma_\mu$ the $S(t)$ exploration exponent and its standard error; Rank~1+2 the combined visitation share of the two most-visited locations; $P_{95}(k)$ the 95th percentile of $R_g$ for users whose $k$-radius of gyration is near zero; $\Delta\sigma_x$/$\Delta\sigma_y$ the percent change in Gonzalez-frame spread relative to the pre-restriction period; and $\hat H$ the peak (modal) value of each entropy measure. The same qualitative pattern reported for the standard 15 March boundary holds under this shifted starting date.}
    \label{tab:supp_dates}
\end{table}
}

{
\renewcommand{\arraystretch}{1.3}
\setlength{\tabcolsep}{8pt}
\begin{table}[t!]
    \centering
    \begin{tabular}{l l l r r r r}
        \hline
        Region & $t^{burst}$ & Comparison & $D_{\text{random}}$ & $D_{\text{uncorrelated}}$ & $D_{\text{real}}$ & $D_{\text{real/random}}$ \\
        \hline
        \multirow{3}{*}{California} & \multirow{3}{*}{1 h}
            & pre -- during  & 0.4494 & 0.3636 & 0.3227 & 0.1278 \\
        &   & pre -- post    & 0.1375 & 0.0484 & 0.0433 & 0.0899 \\
        &   & during -- post & 0.4506 & 0.3421 & 0.3073 & 0.1199 \\
        \hline
        \multirow{3}{*}{Massachusetts} & \multirow{3}{*}{1 h}
            & pre -- during  & 0.4662 & 0.4066 & 0.3626 & 0.1353 \\
        &   & pre -- post    & 0.2228 & 0.1072 & 0.0971 & 0.1044 \\
        &   & during -- post & 0.5245 & 0.4298 & 0.3883 & 0.1549 \\
        \hline
        \multirow{3}{*}{California} & \multirow{3}{*}{8 h}
            & pre -- during  & 0.3697 & 0.2659 & 0.2256 & 0.0879 \\
        &   & pre -- post    & 0.2699 & 0.1340 & 0.1296 & 0.0817 \\
        &   & during -- post & 0.4954 & 0.3261 & 0.2913 & 0.0932 \\
        \hline
        \multirow{3}{*}{Massachusetts} & \multirow{3}{*}{8 h}
            & pre -- during  & 0.3911 & 0.3085 & 0.2729 & 0.0867 \\
        &   & pre -- post    & 0.3630 & 0.2015 & 0.1877 & 0.0875 \\
        &   & during -- post & 0.5717 & 0.4018 & 0.3620 & 0.1130 \\
        \hline
        \multirow{3}{*}{California} & \multirow{3}{*}{24 h}
            & pre -- during  & 0.3192 & 0.2848 & 0.2713 & 0.0966 \\
        &   & pre -- post    & 0.3450 & 0.1421 & 0.1269 & 0.2315 \\
        &   & during -- post & 0.5268 & 0.3468 & 0.3210 & 0.1525 \\
        \hline
        \multirow{3}{*}{Massachusetts} & \multirow{3}{*}{24 h}
            & pre -- during  & 0.3538 & 0.3395 & 0.3267 & 0.1301 \\
        &   & pre -- post    & 0.4055 & 0.1891 & 0.1656 & 0.2617 \\
        &   & during -- post & 0.5865 & 0.4257 & 0.3961 & 0.1646 \\
        \hline
    \end{tabular}
    \caption{{\bf Two-sample Kolmogorov-Smirnov statistics for entropy distributions across pandemic phases.} $D$ denotes the KS test statistic comparing the empirical distributions of each entropy measure between two periods (pre-restriction: 15 January -- 15 March; during restriction: 15 March -- 15 May; post-restriction: 15 May -- September). All comparisons are significant at $p < 0.001$ (sample sizes range from $\sim$233,000 to $\sim$1,059,000 per period; see main text, Table~\ref{tabPeopleCal}, and Supplementary Material, Table~\ref{tabPeopleMA}). For the pre--during and during--post comparisons, $D_{\text{real/random}}$ --- the statistic for the real-to-random entropy ratio --- is consistently the smallest of the four measures across every region and burst-filter threshold, often by a factor of 2--5 relative to $D_{\text{random}}$, indicating that the distributional shift associated with the onset and easing of restrictions is driven primarily by the reduction in the number of visited locations $n$ rather than by a change in the sequential structure of trajectories. This pattern does not hold for the pre--post comparison, where $D_{\text{real/random}}$ is comparable to or larger than $D_{\text{real}}$ (most notably at $t^{burst}=24$h, both regions), suggesting a level of noise that hides the otherwise found information.}
    \label{tab:ks_entropy}
\end{table}
}

\end{document}